\documentclass[%
 reprint,
 amsmath,amssymb,
 aps,
]{revtex4-1}
\usepackage{appendix}
\usepackage{graphicx}
\usepackage{dcolumn}
\usepackage{bm}
\usepackage{tabularx}
\usepackage{setspace}

\begin{document}
\renewcommand{\arraystretch}{1.5}

\title{Statistical Mechanics of Dislocation Pileups in Two Dimensions}

\author{Grace H. Zhang}
\affiliation{Department of Physics, Harvard University, Cambridge, MA 02138, USA.}
\author{David R. Nelson}
\affiliation{Department of Physics, Harvard University, Cambridge, MA 02138, USA.}%

\date{\today}

\begin{abstract}
Dislocation pileups directly impact the material properties of crystalline solids through the arrangement and collective motion of interacting dislocations. We study the statistical mechanics of these ordered defect structures embedded in two dimensional crystals, where the dislocations themselves form one-dimensional lattices. In particular, pileups exemplify a new class of inhomogeneous crystals characterized by spatially varying lattice spacings. By analytically formulating key statistical quantities and comparing our theory to numerical experiments using an intriguing mapping of dislocation positions onto the eigenvalues of recently studied random matrix ensembles, we uncover two types of one-dimensional phase transitions in dislocation pileups: a thermal depinning transition out of long-range translational order from the pinned-defect phase, due to a periodic Peierls potential, to a floating-defect state, and finally the melting out of a quasi-long range ordered floating defect-solid phase to a defect-liquid. We also find the set of transition temperatures at which these transitions can be directly observed through the one-dimensional structure factor, where the delta function Bragg peaks, at the pinned-defect to floating-defect transition, broaden into algebraically diverging Bragg peaks, which then sequentially disappear as one approaches the two-dimensional melting transition of the host crystal. We calculate a set of temperature-dependent critical exponents for the structure factor and radial distribution function, and obtain their exact forms for both uniform and inhomogeneous pileups using random matrix theory. 
\end{abstract}

\maketitle

\section{\label{sec:level1}Introduction}

The structure and motion of dislocation assemblies directly alter the mechanical response of crystalline materials. 
Although perfect single crystals and isolated defects (e.g. a single dislocation, a point-like interstitial or vacancy) have been well characterized, how defects behave in organized substructures is less understood~\cite{lesar2014simulations}. Dislocation pileups permeate plastically deformed materials, and are among the most prevalent types of dislocation substructures and the building block of more complex assemblies such as dislocation cell walls~\cite{sethna2017deformation}. 

\begin{figure}[h]
\centering
\includegraphics[width=0.9\columnwidth]{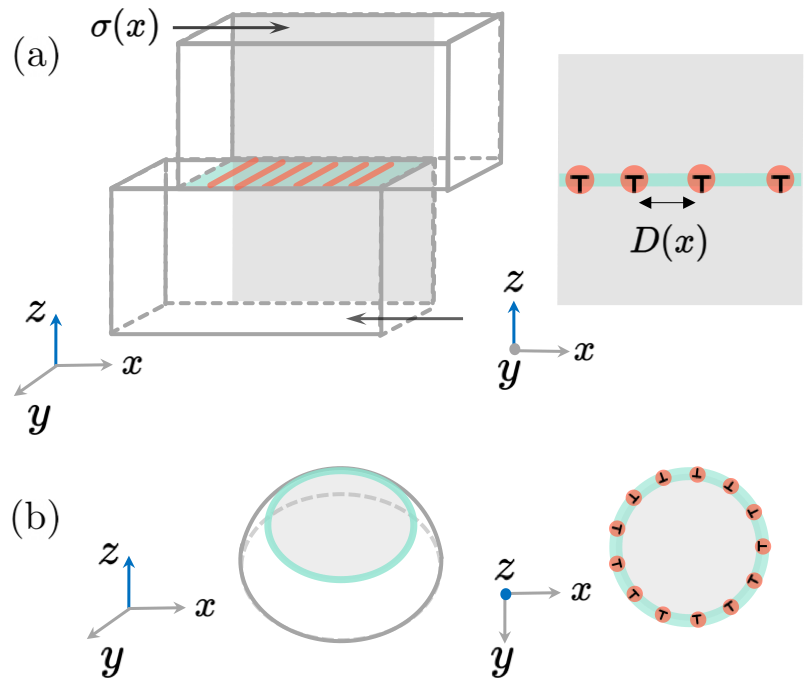}
\caption{ Schematic of conditions under which one-dimensional dislocation pileups, consisting of (orange) short edge dislocation lines (these become point-like when the $y$ dimension is extremely narrow) with Burgers vectors aligned with a glide direction (turquoise strips), embedded in a two-dimensional surface (gray). (a) When a thin two-dimensional crystalline slab experiences applied shear stress $\sigma(x)$, short dislocation lines pile up along the direction of the shear stress. The spatial profile of the applied stress $\sigma(x)$ directly determines the density distribution of the dislocations via Eq.~(\ref{eq:force_balance}). (b): When a two-dimensional crystal is curved, residual stresses due to the Gaussian curvature leads to dislocation pileups that wrap around the spherical cap. Schematic in (b) adapted from Refs.~\cite{azadi2014emergent,azadi2016neutral}). \label{fig:intro_schm}}
\end{figure}

We study here the statistical mechanics of dislocation pileups embedded in two-dimensional (2d) slices of host crystals (see Fig.~\ref{fig:intro_schm}), where defect structures can emerge through shear stress loading, polygonization, and residual stress when the host crystal exhibits nonzero curvature. In 2d host crystals, dislocation pileups exist as one-dimensional (1d) queues of edge dislocations aligned in the same glide plane with Burgers vectors of identical magnitude $b$. Such dislocation arrays are remarkable because of the strong repulsive interactions --- they are only energetically stable if we forbid climb motion out of the glide plane~\cite{hirth1983theory}. They are distinct from other defect structures, such as the Abrikosov flux lattice~\cite{abrikosov2004nobel} and the domain walls that characterize the commensurate-incommensurate transition~\cite{coppersmith1982dislocations}, in that they exemplify a new class of inhomogeneous crystals, with a set of lattice constants $D(x)$ that vary smoothly in space (Fig.~\ref{fig:intro_schm}a). Understanding the statistical mechanics of pileups thus helps elucidate the more general physics associated with higher-dimensional inhomogeneous crystals, which describe a wide variety of systems including plasmas~\cite{thomas1994plasma,mughal2007topological,thomas1996melting}, foams~\cite{drenckhan2004demonstration}, ionic gases~\cite{mielenz2013trapping}, and colloidal particles~\cite{soni2018emergent}. 

\begin{figure}[b]
\centering
\includegraphics[width=1\columnwidth]{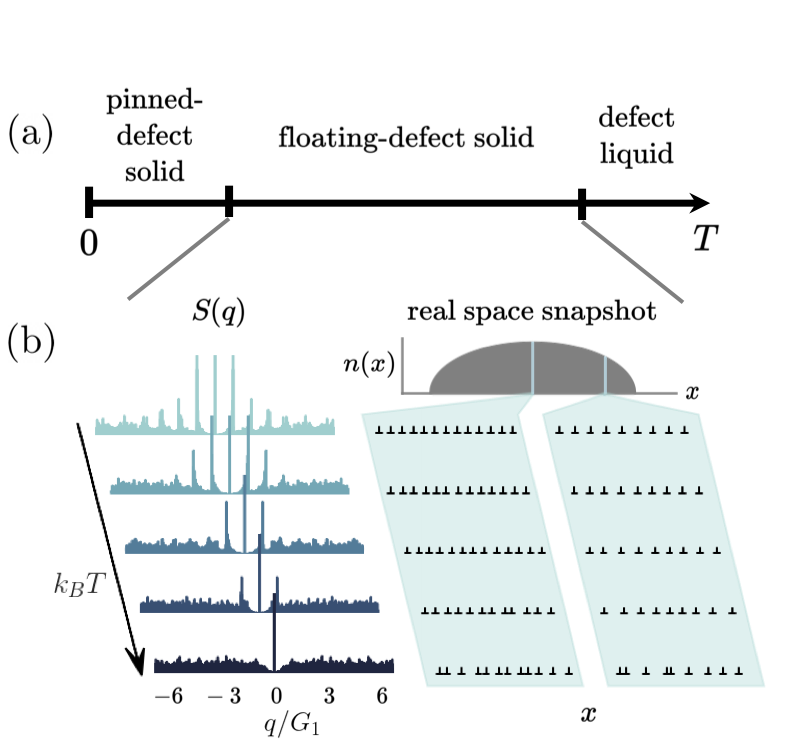}
\caption{(a): Phase diagram for dislocation pileups as a function of temperature $T$. A pinned defect crystal with Delta function Bragg peaks appears at low temperatures, with a floating-defect solid phase that gradually melts at intermediate temperatures. (b): Melting of a semicircular dislocation pileup in the floating-defect solid phase as revealed by random matrix simulations of a semicircular density distribution of defects. Black downward arrow on the left side indicates the direction of increasing temperature $T$. At each temperature, the structure factor $S(q)$ extracted from one random matrix simulation of $N=5000$ total dislocations is shown on the left, and snapshots of the dislocations (with positions given by the eigenvalues of a random matrix) near the lattice center and the dislocations closer to the lattice edge (as indicated in the top schematic) are shown on the right. Here, $n(x)$ denotes the 1d dislocation density profile and $G_1$ is the location of the first Bragg peak. \label{fig:pd}}
\end{figure}

Our results are summarized by the phase diagram in Fig.~\ref{fig:pd}. We map a continuum model of dislocation pileups onto a 1d Coulomb gas of like-magnitude charges, and consider effects of the host lattice by mapping our problem onto a model of quantum Brownian motion~\cite{fisher1985quantum}. By analytically formulating key statistical quantities and renormalization group recursion relations, and numerically testing our theory using mathematical connections with random matrix theory, we uncover a series of one-dimensional defect phase transitions as a function of temperature. 

We first identify an intermediate floating-defect solid phase which exhibits quasi-long range order in one dimension. Upon \textit{increasing} the temperature in the floating-defect solid phase, we identify a remarkable defect melting phase transition to the disordered defect-liquid at finite temperature. This transition proceeds sequentially, as power law divergences at ever-smaller Bragg peaks $\{ G_m \}$ are eliminated, until only the final peak at $G_1$ remains. Upon \textit{decreasing} the temperature, we discover a floating-defect to pinned-defect transition, where translational correlations in dislocation positions transform from quasi-long range order to true long range order. A similar depinning transition was found by Kolomeisky and Straley in a model of a zipper-like interface between two crystalline solids~\cite{kolomeisky1996phase}. There, however, thermal excitations produce approximately equal numbers of oppositely signed dislocation charges. 

While topological defects such as dislocations and disclinations can be crucial in mediating melting transitions of 2d solids~\cite{kosterlitz2017nobel}, we find in our defect melting transition that the dislocations themselves actually undergo a rare case of 1d phase transition out of quasi-long range order, somewhat similar to the melting transition of  2d Abrikosov flux lattices in Type II superconductors~\cite{nelson2002defects}. 
Similarly, while it is known that 2d monolayers adsorbed onto periodic substrates can undergo an incommensurate-commensurate transition~\cite{nelson2002defects} and that dislocations can be pinned by material impurities~\cite{sethna2017deformation}, here we find that the defects can be trapped by the periodic Peierls potential embodied in their own 2d host lattice at sufficiently low temperatures. The intermediate ``floating defect-solid'' we find in one dimension is reminiscient of the 2d floating solid phase hypothesized for 2d monolayers absorbed onto periodic substrates in Refs.~\cite{nelson1979dislocation,halperin1978theory}. 

Both transitions described above can be directly probed through the structure factor $S(q)$. Recall that the energy cost of fluctuations for \textit{short range} interacting particles in 1d is~\cite{kardar2007statistical}
\begin{eqnarray} \label{eq:E_SR}
\Delta E[u(q)]= \frac{1}{2} \int dq~B_0 q^2 |u(q)|^2,
\end{eqnarray}
where $\{u(q)\}$ are the Fourier modes of particle displacements and the bulk modulus $B_0$ is a constant in the long wavelength hydrodynamic limit. It is straightforward to show that the structure factor of short range interacting particles in 1d then exhibits finite Gaussian Bragg peaks at the reciprocal lattice vectors $G_m =  m 2 \pi/D$, where $D$ is the lattice spacing, indicating real-space correlations that decay exponentially with distance (see Appendix~\ref{app:SR} for details). For defect crystals such as pileups, however, different physics emerges from the \textit{long range} interactions between the dislocations, leading to different energetics at small wavevectors. Specifically, the constant $B_0$ in Eq.~(\ref{eq:E_SR}) becomes inversely proportional to the wave vector,
\begin{eqnarray} \label{eq:Bq_intro}
B_0 \rightarrow B(q) = \frac{Yb^2}{8 \pi D^2} \frac{1}{|q|},
\end{eqnarray}
which drastically alters the long wavelength physics of 1d dislocation assemblies ($Y$ is the 2d Young's modulus of the host crystal and $b$ is the magnitude of the Burgers vector characterizing the dislocations). 

When the pileup is in the floating-defect solid phase, the structure factor $S(q)$ exhibits algebraically diverging Bragg peaks, where each Bragg peak at $q = G_m = \frac{2 \pi}{D} m$, $m = 1, 2, \cdots,$  has a distinct temperature-dependent critical exponent $1 - \alpha_m(T)$,
\begin{eqnarray}
\lim_{q \rightarrow G_m} S(q) \sim \frac{1}{|q - G_m|^{ 1 - \alpha_m(T)}},
\end{eqnarray}
with
\begin{eqnarray}
\alpha_m(T) = m^2 \frac{16 \pi k_B T}{Yb^2}. 
\end{eqnarray}
As the temperature increases, the Bragg peaks at ${q=G_m}$ disappear sequentially at temperatures above $T_c^{(m)}$, where 
\begin{eqnarray} \label{eq:Tcm}
k_B T_c^{(m)} = \frac{1}{m^2} \frac{Yb^2}{16 \pi}, \quad m = 1, 2, \cdots, 
\end{eqnarray}
with the higher order Bragg peaks \textit{further} away from the origin in momentum space $q=0$ remaining finite about a \textit{lower} transition temperature. Remarkably, the spacing $D$ between the dislocations in the pileup drops out of this formula. Around the temperature at which the last remaining Bragg peak ($m=1$) disappears, the 2d host crystal (provided it does not melt earlier due to a first order transition) will also melt due to a dislocation-unbinding transition (see discussion in Sec.~\ref{sec:PT})~\cite{kosterlitz1973ordering,nelson1979dislocation,halperin1978theory}. Thus, as the temperature reaches $k_B T_c^{(1)}$, the 1d pileup melts together with the 2d host solid. The transition at lower temperatures, e.g. $k_B T_c^{(2)} = \frac{1}{4} \frac{Yb^2}{16 \pi}$ might be easier to observe experimentally. 

At temperatures $T$ below a characteristic pinning temperature, $T < T_P$, pileups transition to the pinned-defect phase, where quasi-long range order gives way to true long range order, and the algebraically {diverging} Bragg peaks transform into delta function Bragg peaks. We expect that the pinning transition temperature $T_P$ is bounded from above by the pinning temperature associated with a nearby ``accidentally commensurate'' dislocation density:
\begin{eqnarray}
k_B T_P^0 = \frac{2}{M^2} \frac{Yb^2}{16 \pi},
\end{eqnarray}
where $M $ is an integer associated with a commensurate dislocation spacing $M = \frac{D}{a}$ and $a$ is the host crystal lattice constant. 
Since the dislocation spacing is typically much larger than the host lattice constant, $M \gg 1$, the pinning temperature is significantly lower than the melting temperatures associated with the Bragg peaks of the floating solid phase, e.g., $T_P\leq T_P^0 \ll T_c^{(2)} = \frac{1}{4} T_m$, where $T_m = \frac{Yb^2}{16 \pi}$ is the Kosterlitz-Thouless melting temperature of the host crystal~\cite{nelson1979dislocation,halperin1978theory,kardar2007statistical,kosterlitz2017nobel}.  A simplified model for the statistical mechanics of symmetric low angle grain
boundaries (LAGBs), another type of one-dimensional dislocation assembly (with Burgers vectors aligned on average perpendicular instead of parallel to the wall, as for pileups), was studied in Ref.~\cite{zhang2020LAGB}. Some aspects of the commensurate-incommensurate/pinning transitions studied here for inhomogeneous pileups, leading to delta function Bragg peaks in defect structure functions at low temperatures, might be relevant for this problem as well. However, we expect that any modulating potential along the grain for LAGBs is much weaker and more inhomogeneous than the Peierls potential for transverse dislocation glide motions studied in this paper. 

In Sec.~\ref{sec:DP}, we review the continuum theory of one-dimensional dislocation pileups, consisting of edge dislocations confined to the glide plane of a two-dimensional host crystal, and introduce the random matrix models onto which two types of dislocation pileups we discuss can be exactly mapped. In Sec.~\ref{sec:PT}, we identify the melting transitions from quasi-long range order in a floating solid phase through the sequential disappearance of algebraically diverging Bragg peaks in the defect structure factor $S(q)$. We first establish the theory for uniform pileups with equally spaced dislocations, and subsequently for inhomogeneous pileups. We also examine the radial distribution function $g(r)$ and find that correlations as a function of  inter-dislocation distance decay with a power law envelope, described by another temperature-dependent critical exponent related to $\alpha_1(T)$. We then check our predictions numerically using random matrix simulations (RMS). These random matrix simulations have a continuously tuneable temperature parameter, and are highly efficient compared to conventional molecular dynamics or Monte Carlo simulations of long range interactions. In Sec.~\ref{sec:host}, we study the effect of a periodic Peierls potential on pileups, and identify the transition from the floating-defect phase to a low-temperature pinned-defect phase by mapping our problem onto a quantum Brownian motion model~\cite{fisher1985quantum}, which we analyze by deriving the renormalization group recursion relations. 

\section{Dislocation Pileups and Random Matrix Theory \label{sec:DP}}
In this section, we review the equilibrium properties of dislocation pileups and show that the statistical mechanics of two types of pileups can be mapped exactly onto the eigenvalue statistics of recently-studied random matrix ensembles~\cite{dumitriu2002matrix}. Table~\ref{tab:disloc} summarizes the equilibrium dislocation densities and random matrix connections (if they exist) for the pileups studied in this work. 

Dislocation pileups form in crystals under applied shear stress $\sigma$. Although this macroscopic shear stress is often taken to be a constant, we have found it convenient to allow it to depend on position $x$, $\sigma = \sigma(x)$, which allows us to study a broader class of pileups. As we will show, a pileup in the floating-defect phase embedded in a 2d crystal behaves like a Coulomb gas of like-signed charges with logarithmic interactions confined to one dimension. Although dislocations with like-signed Burgers vectors in the same glide plane tend to expel each other outwards indefinitely, dislocations in a pileup are confined by physical obstacles and/or grain boundaries, or equivalently by external potentials generated by applied shear stresses $\sigma(x)$ that are spatially non-uniform~\cite{chakravarthy2011stress,gao1999mechanism}. As discussed below, the force balance condition from a continuum model determines the (possibly inhomogeneous) equilibrium dislocation density of each pileup in response to the applied shear stress. 

Pileups can also occur on \textit{curved} 2d crystals in response to curvature-induced residual stress projected onto the glide plane~\cite{azadi2014emergent,azadi2016neutral}. For example, edge dislocations on a spherical cap can form pileups along the latitudinal direction near the cap boundary (see illustration in Fig.~\ref{fig:intro_schm}b). 

\subsection{Equilibrium dislocation densities of pileups}

The Hamiltonian for a one-dimensional dislocation pileup embedded in a two-dimensional host crystal is~\cite{hirth1983theory}
\begin{eqnarray} \label{eq:H_cont_general}
H [ n ( x ) ] =&& \int_{-L/2}^{L/2} d x~ n ( x )  b \sigma_0 U(x) \\
 - \frac{1}{2} &&\frac{Yb^2}{4 \pi}\int_{-L/2}^{L/2} dx \int_{-L/2}^{L/2} dx^\prime n ( x ) n \left( x ^ { \prime } \right) \ln{| x - x ^ { \prime } |}, \notag
\end{eqnarray}
where $Y$ is the 2d Young's modulus, $b$ is the magnitude of the Burgers vector, and $n(x)$ is the density of dislocations along the pileup. (Eventually, we will take the discrete dislocation positions $\{x_n \}$ into account by setting $n(x) = \sum_n \delta(x-x_n)$, but here it is convenient to use a more general continuum notations.) The first term, with ${\sigma_0 U(x) = \int_{-L/2}^x dx' \sigma(x')}$ comes from the Peach-Koehler force due to the applied shear stress $\sigma(x)$~\cite{peach1950forces}, where $\sigma_0$ measures the strength of the shear stress and $U(x)$, with dimensions of length, is the spatial profile of the potential experienced by the dislocations due to the shear stress. Note that the sign of the dislocation density $n(x)$ indicates the direction of the local Burgers vector $\vec b = \pm b \hat x$, directed along the pileup. With the exception of the double pileup (first row of Table~\ref{tab:disloc}), all pileups studied here have edge dislocations with Burgers vectors of the same sign. 

The average dislocation density $n(x)$ can be calculated from the applied stress ($\sim \partial_x U(x)$) via the force balance condition at equilibrium, obtained from Eq.~(\ref{eq:H_cont_general}) by a functional derivative with respect to $n(x)$ followed by a spatial derivative with respect to $x$,
\begin{eqnarray} \label{eq:force_balance}
0 =\sigma_0 b \frac{d U(x)}{dx} +\frac{Yb^2}{4 \pi}\int_{-L/2}^{L/2} dx^\prime \frac{ n \left( x ^ { \prime } \right)}{x' - x  }.
\end{eqnarray}
Eq.~(\ref{eq:force_balance}) can be solved for many interesting cases using special solutions to the Hilbert transform, given by the Tschebyscheff (Chebyshev) polynomials~\cite{hirth1983theory}. Thus, by varying the form of the profile $U(x)$ through the applied stress, one can obtain an entire class of inhomogeneous dislocation pileups in one dimension, each with its own distinctive density profile. 

We summarize the density distributions for the different pileups studied in this paper in Table~\ref{tab:disloc}, and refer the readers to Appendix~\ref{app:DP} for details of the derivations using the framework described above. Table~\ref{tab:disloc} reveals the rich variety of pileups possible depending on the spatial profile of the shear stress $\sigma(x) = \sigma_0 \partial_x U(x)$. Double pileups and single pileups experience uniform stress fields $\sigma(x) = \sigma_0$ and linear potentials $U(x) \sim x$, while semicircle pileups result from linearly varying stress fields $\sigma(x) \sim x$, corresponding to a quadratic confining potential $U(x) \sim x^2$. 

While the statistical mechanics ideas used here apply generally to any one-dimensional pileup embedded in a two-dimensional crystal, we will utilize the specific pileups in Table~\ref{tab:disloc} to explicitly demonstrate and check various aspects of our theory. The theory of \textit{uniform} pileups is an important building block for understanding inhomogeneous pileups. A pileup with uniform average density can be constructed in two ways. A uniform pileup ring follows from imposing periodic boundary conditions (Row 5 of Table~\ref{tab:disloc}) without a confining potential, while a uniform pileup chain with open boundary conditions requires a non-uniform central potential profile $U(x) = U_U(x)$. The form of $U_U(x)$ follows from inverting the definition given for an average density $n(x) = n_U(x)$ described by a rectangle function
\begin{eqnarray}
n_U(x) = n_\text{U} \Pi\left( \frac{x}{L} \right) \equiv \begin{cases} 1 & \left| \frac{x}{L} \right| < \frac{1}{2} \\ 
\frac{1}{2} & \left| \frac{x}{L} \right| = \frac{1}{2}  \\
0 & \left| \frac{x}{L} \right| > \frac{1}{2}  \end{cases}
\end{eqnarray}
where $n_\text{U}$ is a constant and $\Pi(z)$ is the rectangle function with the following well-defined Hilbert transform~\cite{bracewell1986fourier},
\begin{eqnarray}
\mathcal{H}[\Pi(z)]&=&\frac{1}{\pi} \text{PV} \int_{-\infty}^{\infty} \frac{\Pi(z) d z}{z-y} \\
&=&\frac{1}{\pi} \text{PV} \int_{-1/2}^{1/2} \frac{\Pi(z) d z}{z-y} = \frac{1}{\pi} \ln \left|\frac{y-\frac{1}{2}}{y+\frac{1}{2}}\right|,
\end{eqnarray}
where PV denotes the principal value of the integral. Upon integrating the result of the Hilbert transform, the central potential for a \textit{uniform} pileup chain is obtained as 
\begin{eqnarray} \label{eq:U_uni_main}
U_\text{U}(x) &=& \left( \frac{L}{2} - x \right) \ln \left(  \frac{L}{2} - x \right) \\
&&+  \left( \frac{L}{2} + x \right) \ln \left( \frac{L}{2} + x \right) \notag
\end{eqnarray}
(see row 4 of Table~\ref{tab:disloc} and Appendix~\ref{app:DP} for details). Note that the dislocation density described by the rectangle function breaks translational invariance of the pileup and the confining potential $U_\text{U}(x)$ diverges at $|x| = L/2$, corresponding physically to impenetrable walls bounding a row of like-signed dislocations. 

Interestingly, the central potential for a uniform pileup $U_\text{U}(x)$ varies quadratically near the pileup center, just like the central potential $U_\text{SC}(x)$ for a semicircular pileup (a pileup with semicircular density distribution, see row 3 of Table~\ref{tab:disloc}), and the dislocation density near the center of the semicircular pileup $n_\text{SC}(x)$ is close to uniform, as for the rectangle density $n_\text{U}(x)$:
\begin{eqnarray}
\lim_{x \rightarrow 0} n_\text{SC}(x) &\sim& n_\text{U}(x) \label{eq:SC_U_n} \\
\lim_{x \rightarrow 0} U_\text{U}(x) &\sim& U_\text{SC} (x). \label{eq:SC_U_sigma}
\end{eqnarray}

Remarkably, as shown in the next section, the statistical mechanics of semicircular pileups ($U(x) \sim x^2$) and single pileups ($U(x) \sim x$, row 2 of Table~\ref{tab:disloc}) at finite temperatures maps exactly on to the eigenvalue statistics of special random matrix ensembles that are easy to simulate. Random matrix simulations then allow us to check our theoretical predictions in Sec.~\ref{sec:PT}. In particular, while a semicircular pileup is inhomogeneous when considered in its entirety, we can use its center region to approximate a uniform pileup. In Sec.~\ref{sec:PT}, we utilize this feature to numerically check our theoretical structure functions and radial distribution functions for uniform pileups and inhomogeneous semicircular pileups. 

\small 
\begin{table*}[t]
\begin{tabular}{@{}c|c|c|c@{}}
\toprule
\multicolumn{1}{c|}{Dislocation Pileup}                                     & \multicolumn{1}{c|}{Schematic} & \multicolumn{1}{c|}{\begin{tabular}[c]{@{}c@{}}Dislocation density $n(x)$ \\ $x \in \left (-\frac{L}{2}, \frac{L}{2} \right)$\end{tabular}
}                                                                                                & \begin{tabular}[c]{@{}c@{}}Central potential profile $U(x)$ \\ $x \in \left (-\frac{L}{2}, \frac{L}{2} \right)$\end{tabular}  \\ \hline
Double pileup                                                               &  
\raisebox{-.4\height}{\includegraphics[width=0.18\columnwidth]{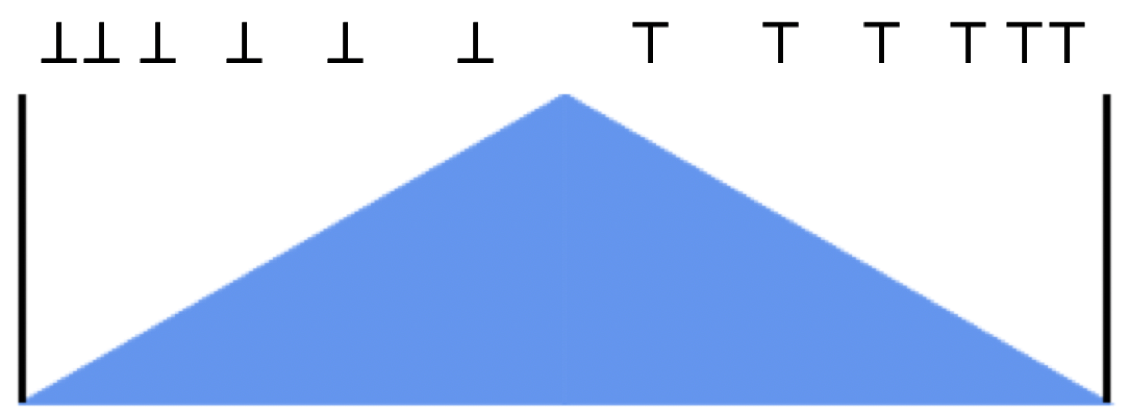}}
& \begin{tabular}[c]{@{}c@{}}$n_\text{D}(x)=\zeta \frac{x}{\sqrt{\left(\frac{L}{2}\right)^2-x^2}}$\\ $N_\text{D} = \zeta L$\end{tabular}                  & $- |x|$  \\ \hline
\begin{tabular}[c]{@{}c@{}}Single pileup \\ (RM: Wishart)\end{tabular}      &
\raisebox{-.4\height}{\includegraphics[width=0.18\columnwidth]{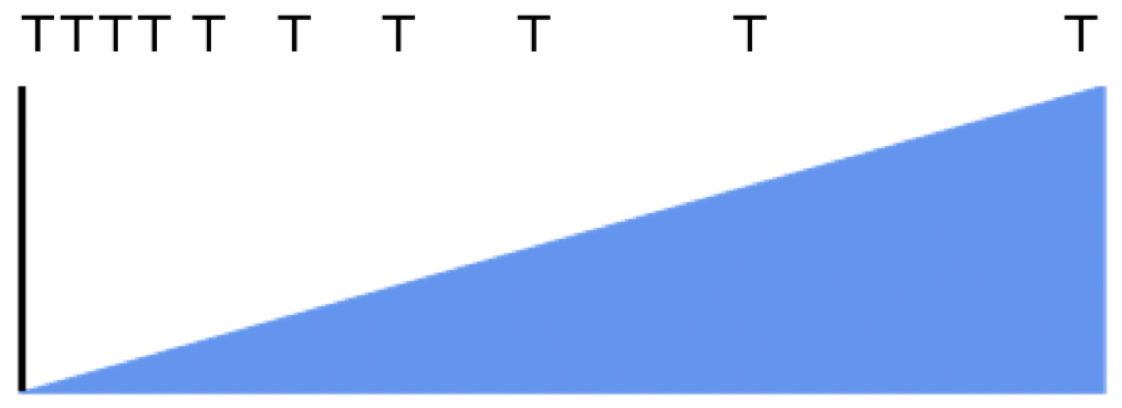}}
& \begin{tabular}[c]{@{}c@{}}$n_\text{S}(x)=\zeta \frac{\sqrt{\frac{L}{2}- x}}{\sqrt{\frac{L}{2}+x}}$\\ $N_\text{S} = \zeta L \frac{\pi}{2}$\end{tabular} & $x$   \\ \hline
\begin{tabular}[c]{@{}c@{}}Semicircle lattice\\ (RM: Gaussian)\end{tabular} &
\raisebox{-.4\height}{\includegraphics[width=0.18\columnwidth]{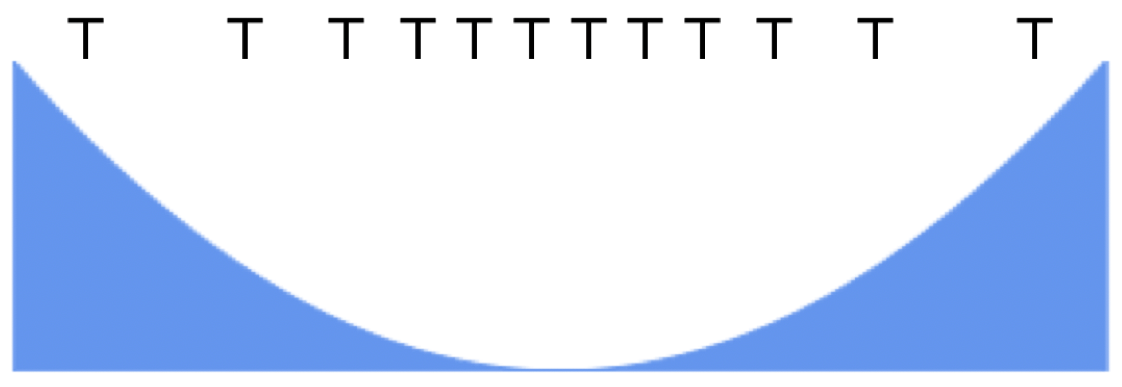}}& \begin{tabular}[c]{@{}c@{}}$n_\text{SC}(x)=\zeta \sqrt{1-\left( \frac{x}{L/2} \right)^2}$\\ $N_\text{SC} = \zeta_{} L \frac{\pi}{4}$\end{tabular}      & $\frac{1}{L / 2} \frac{x^{2}}{2}$   \\ \hline
Uniform lattice                                                             &   \raisebox{-.4\height}{\includegraphics[width=0.18\columnwidth]{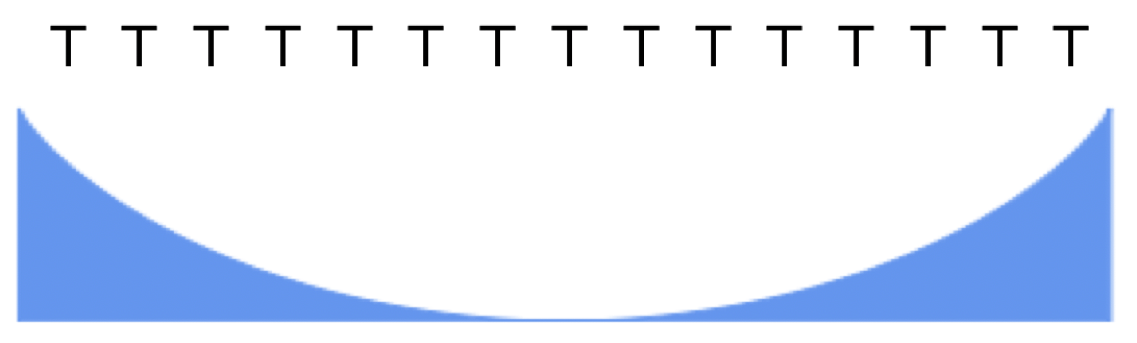} }                            & \begin{tabular}[c]{@{}c@{}}$n_\text{U}(x)=\zeta \pi$\\ $N_\text{U} = \zeta L \pi$\end{tabular}                                                          &  \begin{tabular}[c]{@{}c@{}}$\left(\frac{L}{2}-x\right) \log \left(\frac{L}{2}-x\right)$\\ $\quad +\left(x+\frac{L}{2}\right) \log \left(x+\frac{L}{2}\right)$\end{tabular}   \\ 
\hline
Uniform ring                                                             &   \raisebox{-.4\height}{\includegraphics[width=0.1\columnwidth]{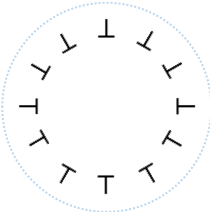} }                            & \begin{tabular}[c]{@{}c@{}}$n_\text{UR}(x)=\zeta \pi$\\ $N_\text{UR} = \zeta L \pi$\end{tabular}                                                          & None \\ \hline
\end{tabular}
\caption{\label{tab:disloc} A summary of the different dislocation pileups discussed in this paper. The ``Schematic'' column shows a typical stationary dislocation distribution and the corresponding confining potential (in blue). In the expressions for dislocation density $n(x)$ and total number of dislocations $N$ in the next column, $\zeta = \frac{4 \sigma_0}{Y b}$, where $Y$ is the 2d Young's modulus, $b$ is the magnitude of the Burgers vector and $\sigma_0$ determines the magnitude of the applied shear stress. There exist two random matrix (RM) ensembles whose eigenvalue statistics correspond exactly to the statistical mechanics of the single pileup ($\beta$-Wishart ensemble) and the semicircular pileup ($\beta$-Gaussian ensemble).  }
\end{table*}
\normalsize 

\subsection{Connection to random matrix theory \label{sec:RMT}}

We introduce two random matrix ensembles that allow efficient finite temperature simulations of the long range interactions embodied in dislocation pileups---the general $\beta$-Gaussian (Hermite) ensemble and general $\beta$-Wishart (Laguerre) ensemble~\cite{dumitriu2002matrix}---whose eigenvalue statistics map exactly onto the statistical mechanics of  semicircular pileups and single pileups. Specifically, the random matrix parameter $\beta$ is proportional to the inverse temperature $1/k_B T$ of the dislocation pileups
\begin{equation} \label{eq:betas}
\beta = \frac{Yb^2}{4 \pi} \frac{1}{k_B T},
\end{equation}
and the joint probability distribution function (JPDF) of the random matrix eigenvalues at a particular value of $\beta$ is equal to the Boltzmann factor (normalized by the partition function) of pileup configurations at the temperature $T$ corresponding to Eq.~(\ref{eq:betas}). (Note that we do \textit{not} set $\beta = 1/k_B T$, the usual notational convention in statistical mechanics.)
Thus, the eigenvalues of these random matrices are the dislocation positions in a snapshot of the pileup in thermal equilibrium, and the temperature at which the snapshot is taken can be tuned via the matrix parameter $\beta$ in Eq.~(\ref{eq:betas}), also known as the random matrix inverse temperature or the Dyson index. 

Importantly, the general $\beta$-Gaussian and the general $\beta$-Wishart random matrices are tridiagonal and allow $\beta$ to assume any positive value $\beta > 0$.
We can thus obtain an equilibrium configuration of semicircular pileups and single pileups at any temperature $k_B T$ by diagonalizing a tridiagonal random matrix, an operation that scales with the total particle number $N$ as $O(N\log(N))$~\cite{coakley2013fast}. Thus, the use of random matrix ensembles allows us to bypass the challenges of direct numerical simulations with, say, molecular dynamics for $N$ particles with long range interactions, which scales as $O(N^{5/2})$~\footnote{Under long range interaction, all $N$ particles must be updated for each of the $N$ equations of motion, scaling as $\sim N^2 \tau $. The time scale $\tau$ is inversely proportional to the longest wavelength modes $\tau^{-1} \sim \omega(q) \sim q^{1/2}  > \sim N^{-1/2}$, giving the total scaling of $O(N^{5/2})$.}

In contrast to the general $\beta$ matrices, the usual classical $\beta$-Gaussian matrices~\cite{mehta2004random} and classical $\beta$-Wishart matrices~\cite{wishart1928generalised}, with eigenvalue statistics identical to their general $\beta$ counterparts, are fully dense and only allow $\beta$ to assume three possible values  $\beta=$1, 2, 4~\cite{livan2018introduction}. 
Nevertheless, the analytical results derived via orthogonal polynomials for these standard random matrix ensembles $(\beta = 1, 2, 4)$~\cite{mehta2004random} will also be useful for us.

\subsubsection{General $\beta$-Gaussian ensemble \label{sec:Gaussian}}

Matrix models of the \textit{general} $\beta$-Gaussian ensembles, where $\beta$ assumes any positive real value, take the following real symmetric \textit{tridiagonal} form \cite{dumitriu2002matrix},
\footnotesize 
\begin{eqnarray}  \label{eq:Hb}
H_\beta = \frac{1}{\sqrt{2}} \left[ \begin{matrix}
N(0, 2)             & \chi_{(N-1) \beta}&                   &     & 0 \\
\chi_{(N-1) \beta} & N(0, 2)           & \chi_{(N-2) \beta}&     & \\
                    & \ddots            & \ddots            & \ddots  & \\
                    &                   & \chi_{2 \beta}    & N(0, 2) & \chi_{ \beta} \\
0                    &                   &                   & \chi_{ \beta}  & N(0, 2) \end{matrix} \right],
\end{eqnarray}\normalsize 
where all elements off the tridiagonals, including the corner entries, are zero. Here, $N(0,2)$ indicates a random number drawn from the normal probability distribution with mean 0 and variance 2, and $\chi_{k}$ represents a random number drawn from the chi distribution, which describes the statistics of $ \sqrt{\sum_{i=1}^k Z_i}$, where $Z_1, \cdots, Z_k$ are $k$ independent normally distributed variables with mean 0 and variance 1. The probability density function $p_k(x)$ corresponding to the chi distribution $\chi_k$ is then~\cite{johnson1995continuous} 
\begin{equation} \label{eq:chi_p}
p_k(x)=\begin{cases}
{\frac{x^{k-1} e^{-x^{2} / 2}}{2^{k / 2-1} \Gamma\left(\frac{k}{2}\right)},} & {x \geq 0} \\
{0,} & {\text { otherwise, }}
\end{cases}
\end{equation}
where $\Gamma(\frac{k}{2})$ is the gamma function, and $k$ does not have to be an integer and can in fact assume any real value. Note that all diagonal elements $H_{\beta, ii}$ are \textit{independently} drawn from $N(0,2)$, whereas each $H_{\beta, ij} = H_{\beta, ji}$ $(i \neq j)$ off-diagonal pair are in fact the same number, so these matrices are symmetric with real eigenvalues. 

Upon rescaling the $N$ eigenvalues $(x_1, \cdots, x_N)$ as {${x_i \rightarrow \sqrt{2 \beta N} x_i}$}, so that the spectrum lies in the interval $x \in (-1, 1)$, the eigenvalue joint probability distribution function (JPDF) is, up to a normalization constant~\cite{dumitriu2002matrix},
\begin{eqnarray} \label{eq:bH_RMT}
n(x_1, \cdots, x_N) \sim e^{ -\beta N \left[ \sum_i x_i^2  - \frac{1}{2 N} \sum_{j\neq k} \ln|x_j - x_k| \right] }. 
\end{eqnarray}
Upon substituting $\beta$ using Eq.~(\ref{eq:betas}) and replacing $N$ using the semicircular pileup normalization condition $N =\pi \sigma_0  L / Yb$  (see row 3 of Table~\ref{tab:disloc}), one can immediately see that the exponential in Eq.~(\ref{eq:bH_RMT}) is exactly equal to the reduced Hamiltonian $H/k_B T$ in Eq.~(\ref{eq:H_cont_general}) for the semicircular pileup in row 3 of Table~\ref{tab:disloc}, which experiences a quadratic central confining potential $U(x) = x^2/L$. 

In the large $N$ limit, the average eigenvalue density distribution  $n(x) = \int dx_2 \cdots dx_N~n(x_1, \cdots, x_N)$ is given by the famous semicircle law~\cite{dumitriu2002matrix,wigner1958distribution},
\begin{equation}
n(x) = \frac{2 N}{\pi } \sqrt{ 1 -x^2}.
\end{equation}
As mentioned previously, the semicircle pileup and the associated general $\beta$-Gaussian random matrices will be exceedingly useful for testing the theory developed in Sec.~\ref{sec:PT}. 

In the next section, we describe another fascinating connection, this time between random matrices and single dislocation pileups. However, single pileups have lattice spacings that are extremely inhomogeneous near the piling edge, so that only a small amount of crystalline order can survive. We will not focus much on them for the remainder of this paper. The reader may skip the next subsection without loss of continuity. 

 \subsubsection{General $\beta$-Wishart ensemble \label{sec:Wishart}}

Matrix models of the general $\beta$-Wishart ensemble consist of square matrices of the form $W_\beta = B_\beta B_\beta^T$, where $B_\beta$ are $N \times N$ square, \textit{bidiagonal}, matrices. The matrix elements contain, in addition to $\beta$, another tuning parameter $M$. (A similar parameter $M$ appears in the classical Gaussian Wishart matrices $W = B B^T$, where $B$ are $N \times M$ rectangular matrices. In generalizing the Wishart ensemble, Ref.~\cite{dumitriu2002matrix} has transformed $M$ from an integer matrix rank parameter into a parameter tuning the probability distribution of the matrix elements.) The subset of these square $B_\beta$ matrices describes the statistical mechanics of single dislocation pileups. The random bidiagonal matrix $B_\beta$ with $M = N$ takes the form
\footnotesize 
\begin{eqnarray} 
B_\beta& =& \left[ \begin{matrix}
\chi_{2 \bar a}             & 0									&       &     & 0\\
\chi_{(N-1) \beta}  		&\chi_{2 \bar a - \beta}           & 0		&     & \\
                    & \ddots            & \ddots            & \ddots  & \\
                    &                   & \chi_{2 \beta}    & \chi_{2\bar a - (N-2)\beta} & 0 \\
0                    &                   &                   & \chi_{ \beta}  & \chi_{2\bar a - (N-1)\beta}  \end{matrix} \right],
\end{eqnarray} \normalsize 
where $\beta$ can assume any positive value, $\bar a = \frac{\beta N}{2}$, and $\chi_k$ indicates a random number drawn from the chi distribution shown in Eq.~(\ref{eq:chi_p}).

Note that the product $W_\beta = B_\beta B_\beta^T $ is a symmetric $N \times N$ square matrix with correlated matrix elements. The eigenvalues $\{x_i\}$ of $W_\beta$, which is a positive semidefinite matrix, are the squares of the singular values $\{\sigma_i\}$ of $B_\beta$: $x_i = \sigma_i^2$. Upon scaling the  $N$ eigenvalues $(x_1, \cdots, x_N)$ of matrix $W_\beta$ according to $x_i \rightarrow \beta N x_i $, one obtains the following spectral JPDF, up to a normalization constant, 
\begin{eqnarray} \label{eq:nexp_single}
n(x_1, \ldots, x_N) &\sim&  e^{-\beta N \left[ \sum_i V(x_i)- \frac{1}{2 N} \sum_{j\neq k} \ln |x_j - x_k|\right]},
\end{eqnarray}
where the associated central potential $V(x)$ is 
\begin{eqnarray} \label{eq:wishart_V_anyN}
V(x) &=& \frac{x}{2}  + \frac{( 2/\beta) -1 }{2N}\ln(x).
\end{eqnarray}

In the thermodynamic limit $N \rightarrow \infty$, the weak logarithmic correction vanishes, and the central potential of the eigenvalues in Eq.~(\ref{eq:wishart_V_anyN}) simplifies, 
\begin{equation}
V(x) =\frac{x}{2}. 
\end{equation}Remarkably, upon substituting $\beta$ using Eq.~(\ref{eq:betas}) and replacing $N$ using the single pileup normalization condition $N =2 \pi \sigma_0  L / Yb$  (row 2 of Table~\ref{tab:disloc}), the thermodynamic limit of the exponential weight in Eq.~(\ref{eq:nexp_single}) corresponds exactly to the Hamiltonian $H/k_B T$ of the single pileup, which exhibits a linear potential (i.e. a constant stress field, see row 2 of Table~\ref{tab:disloc}). The average eigenvalue density $n(x) = \int dx_2 \cdots dx_N~n(x_1, x_2, \cdots, x_N)$ in this large $N$ limit is given by the Marchenko-Pastur law~\cite{dumitriu2002matrix,marvcenko1967distribution}, with
\begin{eqnarray} 
n(x) = \frac{N}{2 \pi} \frac{\sqrt{ 4-x}}{\sqrt{x}}, 
\end{eqnarray}
where the spectral support lies in the interval $x \in (0, 4)$. If we shift $x$ by a constant $x \rightarrow x + 2$, the average density becomes
\begin{eqnarray} \label{eq:rho_wishart}
n(x) = \frac{N}{2 \pi} \frac{\sqrt{ 2-x}}{\sqrt{2+x}}.
\end{eqnarray}
The eigenvalue distribution shown in Eq.~(\ref{eq:rho_wishart}) then corresponds exactly to the dislocation density of a single pileup with length $L = 4$, bounded by an impenetrable wall at $x = -2$ and extending towards the positive $x$ direction (see row 2 of Table~\ref{tab:disloc}), with
\begin{eqnarray} \label{eq:n_single_comp}
n_S(x) =\frac{Yb}{4 \sigma_0} \frac{\sqrt{\frac{L}{2} - x}}{\sqrt{\frac{L}{2} + x}},
\end{eqnarray}
where $x \in \left (-\frac{L}{2} , \frac{L}{2} \right)$. 

Note, however, that this correspondence between the statistical mechanics of single pileups and the eigenvalue statistics of the general $\beta$-Wishart ensemble is only exact in the thermodynamic limit of large matrix rank $N$, whereas the statistical mechanics of semicircular pileups maps exactly onto the eigenvalue statistics of the general $\beta$-Gaussian ensemble for all $N$ (i.e. for an arbitrary number of dislocations or eigenvalues).  

\section{Melting transition of dislocation pileups \label{sec:PT}}

In this section, we calculate the structure factors and spatial correlation functions for thermally excited dislocation pileups. We begin by building the theory for uniform pileups and then extend it to inhomogeneous pileups with slowly varying dislocation spacings. We show that an entire sequence of phase transitions can be associated with power-law divergences at different Bragg peaks in the structure factor and identify a set of transition temperatures. We then efficiently simulate semicircular pileups (i.e. pileups with a semicircular average density profile) by diagonalizing the general $\beta$-Gaussian random matrices introduced in Sec.~\ref{sec:RMT}, and extract the structure factor $S(q)$ and the radial distribution function $g(r)$ for both a truncated piece of homogeneous dislocation lattice and the untruncated semicircular dislocation lattice. The simulation results show excellent agreement with our theory.   

\subsection{Energy of fluctuations}
When dislocations with identical Burgers vectors $b > 0$ are arranged at discrete positions $\{x_n \}$, so that $n(x) = \sum_n \delta (x-x_n)$, the Hamiltonian for a pileup takes the following form
\begin{equation} \label{eq:H_discrete}
H =\sum_n B(x_n) - A \sum_{n \neq m} \ln |x_n - x_m|, 
\end{equation}
where the sums are over all dislocations in the pileup, $B(x) = \sigma_0 b U(x)$ is the central confining potential and $A = \frac{1}{2} \frac{Yb^2}{4 \pi}$ is a constant, depending on the Burgers vector magnitude $b$, the Young's modulus $Y$ of the host lattice, and the strength of the applied force $\sigma_0$. 

The change in energy due to particle displacements from equilibrium~\cite{fisher1979defects} in Eq.~(\ref{eq:H_discrete}) is
\begin{eqnarray}
\Delta E  && = \sum_n B(R_n+ u_n) \\
&& - A \sum_{n \neq m} \Big[  \ln | R_n + u_n - (R_m + u_m) | -\ln | R_n - R_m | \Big], \notag
\end{eqnarray}
where $R_n=nD$ is the equilibrium lattice position of the $n$-th dislocation, and $u_n$ is the displacement of the $n$-th dislocation from $R_n$. We will expand each term to obtain the energy of fluctuations to quadratic order in the displacements $\{u_n \}$. 

The uniform pileup ring with periodic boundary conditions (see last row of Table~\ref{tab:disloc}) does not have a confining potential, 
\begin{eqnarray}
B(x) = 0,
\end{eqnarray}
so the energy cost of fluctuations to quadratic order comes solely from the interaction term:
\begin{eqnarray} \label{eq:DeltaE_ring}
\Delta E_\text{Ring} = \frac{A}{2} \sum_{n \neq m} \frac{ (u_n - u_m)^2 }{(R_n - R_m)^2}. 
\end{eqnarray}

In contrast, the uniform pileup chain experiences the confining potential in Eq.~(\ref{eq:U_uni_main}), and the energy of displacements is, approximately,
\begin{eqnarray} \label{eq:DE}
\Delta E &=&  \sum_n E^{(a)}_n u_n^2 - A\sum_{n \neq m} \frac{u_n u_m }{(R_n - R_m)^2},
\end{eqnarray}
where
\begin{eqnarray} \label{eq:Ena}
E_n^{(a)} = \frac{1}{2} \partial_x^2 B(R_n) + A \sum_m {\vphantom{\sum}}' \frac{1}{(R_n - R_m)^2}. 
\end{eqnarray}
Here, $\sum ' $ indicates a sum of over all lattice sites $m$ except $m=n$. 
We can compare the magnitudes of the central potential term and the interaction potential term in $E^{(a)}_n$ in the large $N$ limit by seeing how these two terms scale with the total number of dislocations $N$. 
With the help of Eq.~(\ref{eq:U_uni_main}),  we obtain the first term in Eq.~(\ref{eq:Ena}) as
\begin{eqnarray} \label{eq:Bpp}
\frac{1}{2} \partial_x^2 B(x) = \frac{2}{L} \frac{\sigma_0 b}{1-\left(\frac{x}{L/2} \right)^2}. 
\end{eqnarray}
The energy cost of fluctuations near the edges $|x| \approx L/2$ of the uniform dislocation chain diverges due to the confining nature of the potential. This result is plausible because, by construction, $U_\text{U}(x)$ constrains the dislocations to a density distribution $n_\text{U}(x) = \Pi(x/L)$ that vanishes for $|x|>L/2$, making it infinitely costly for a dislocation on the edge to fluctuate into the forbidden region. Henceforth, we will focus our attention deep inside the pileup chain, where $\partial^2_x B(x) \sim \sigma_0 b \frac{1}{L}$ from Eq.~(\ref{eq:Bpp}). Then, using the normalization condition for uniform pileups $N = 4 \sigma_0 L \pi/Yb$ (row 4 of Table~\ref{tab:disloc}), we see that $A = Yb^2/8 \pi =\sigma_0 b L/2N$, so the second term in Eq.~(\ref{eq:Ena}) scales as
\begin{eqnarray}
A \sum_{m} {\vphantom{\sum}}' \frac{1}{(R_n - R_m)^2} \approx \frac{\pi^2}{3} A \frac{1}{D^2} \sim A \frac{N^2}{L^2} \sim \sigma_0 b \frac{N}{L}. 
\end{eqnarray}
Inside the uniform pileup, the ratio between the two terms in the diagonal energies $\{E^{(a)}_n \}$ in Eq.~(\ref{eq:Ena}) is then
\begin{eqnarray}
\frac{\partial_x^2 B(x)}{A \sum_{m}' (R_n - R_m)^{-2}}  \approx \frac{3}{\pi^2}\frac{B''(x)}{A D^{-2}} &\sim& \frac{1}{L}\frac{L}{N}  \sim \frac{1}{N}. 
\end{eqnarray}
Thus, in the large $N$ limit, we can ignore the contribution due to the confining potential in the bulk of the pileup, and Eq.~(\ref{eq:DE}) becomes
\begin{eqnarray} \label{eq:DeltaE_chain}
\Delta E \approx \frac{A}{2} \sum_{n \neq m} \frac{ (u_n- u_m)^2 }{(R_n - R_m)^2}. 
\end{eqnarray}
The behavior of the fluctuations deep in the interior of a uniform dislocation pileup chain with a confining potential is thus equivalent to that in the uniform ring in Eq.~(\ref{eq:DeltaE_ring}), which has no confining potential to begin with. Neglecting the effect of the confining potential in the uniform pileup chain is somewhat analogous to ignoring the effect of boundary conditions on the bulk properties in the thermodynamic limit. 

We now write $\Delta E$ in terms of Fourier modes using the following Fourier transform conventions,
\begin{eqnarray}
u(q) &=& D \sum_{n = 1}^{N} e^{i q nD} u_n \\
u_n &=& \int_{-\pi/D}^{\pi/D} \frac{dq}{2 \pi} e^{-iqnD} u(q),
\end{eqnarray}
where $\sum_n$ sums over all $N$ lattice sites. 
Eqs.~(\ref{eq:DeltaE_ring}) and (\ref{eq:DeltaE_chain}) then become
\begin{eqnarray}
\Delta E =&& \sum_{n\neq m}  \frac{A}{(R_n - R_m)^2}  \int \frac{dq}{2 \pi} \int \frac{dq'}{2 \pi} u(q) u(q') \notag \\
&& \times \left( e^{i (q+q') n D} - e^{i q n D} e^{i q' m D} \right) . \label{eq:DE_mid}
\end{eqnarray}
Upon relabeling the equilibrium site variables as $ R_{j} = R_n - R_m$ and $R_{\rho} = (R_n + R_m)/2$ and summing over $R_\rho$, Eq.~(\ref{eq:DE_mid}) becomes
\small 
\begin{eqnarray}
\Delta E &=&  \int \frac{dq}{2 \pi} \frac{2 A}{D}\sum_{j > 0} \frac{1}{(R_j)^2}\left( 1 - \cos(R_j q) \right) |u(q)|^2 \\ 
&=& \int \frac{dq}{2 \pi} \frac{1}{2} B(q) q^2|u(q)|^2, \label{eq:DeltaE}
\end{eqnarray}
\normalsize 
where
\begin{eqnarray} \label{eq:kernel}
B(q) q^2 &=& \frac{4 A}{D}\sum_{j > 0} \frac{1}{(R_j)^2}\left( 1 - \cos(R_j q) \right) \\
 &=& \frac{4 A}{D^3}\sum_{j > 0} \frac{1}{n^2}\left( 1 - \cos(nqD) \right) . 
\end{eqnarray}
With the help of the following summation identity~\cite{moretti2004depinning,chui1983grain},
\begin{eqnarray}
\sum_{n=1}^{\infty} \frac{1-\cos (nq D)}{n^{2}}=\frac{\pi|qD|}{2}-\frac{(qD)^{2}}{4} + \cdots
\end{eqnarray}
we truncate the kernel in Eq.~(\ref{eq:kernel}) to lowest order in $q$, 
\begin{eqnarray} \label{eq:sum_j}
B(q) q^2 = \frac{2 A \pi}{D^2}  |q|,
\end{eqnarray}
which dominates the integral in Eq.~(\ref{eq:DeltaE}). 
The change in the long wavelength energy as a function of particle displacements in momentum space is thus
\begin{eqnarray} \label{eq:deltaE_final}
\Delta E  &=&  \int \frac{dq}{2 \pi} \frac{ A\pi }{D^2}|q||u(q)|^2 = \frac{1}{2} \left( \frac{ Yb^2}{4 D^2} \right)\int \frac{dq}{2 \pi} |q||u(q)|^2. 
\end{eqnarray}
Note that the coefficient of $|u(q)|^2$ is linear in $q$, in contrast to elastic theories with short range interactions, where the elastic energies are quadratic in $q$~(see Appendix~\ref{app:SR} or Ref.~\cite{kardar2007statistical}). This linear dependence on $q$ is critical for obtaining singular phenomena associated with phase transitions at the higher order Bragg peaks in the structure factor for one-dimensional pileups.

\subsection{Structure factor for uniform pileups} \label{sec:sq_calc}

The structure function $S(q)$ measures the sensitivity of a crystal to density perturbations of some length scale $\lambda(q) = 2 \pi/q$. We calculate the dominant contributions to the structure factor in two separate regimes: (1) when the momentum $q$ is near 0, which describes long wavelength density fluctuations, and (2) when $q$ is close to the $m$-th reciprocal lattice vector $G_m = \frac{m 2 \pi}{D}$, where $D$ is the constant dislocation spacing. Understanding these two regimes captures the most important features of the structure factor, as confirmed by random matrix simulations in Sec.~\ref{sec:RMS_uni}.

\subsubsection{Long wavelength limit}

Let us again write the microscopic dislocation density $\rho_\text{micro} (x)$ of a single realization of the uniform pileup as
\begin{eqnarray}\label{eq:rho}
\rho_\text{micro} (x) = \sum_{j=1}^N \delta(x-x_j)
\end{eqnarray}
where $\{ x_j \}$ is the set of $N$ dislocation positions. By averaging Eq.~(\ref{eq:rho}) over a hydrodynamic averaging volume centered at $x$, containing a number of dislocations, we can coarse grain $\rho_\text{micro}(x)$ to obtain a smoothed density field $\rho(x)$~\cite{nelson2002defects}. Density fluctuations can then be expressed as $\delta \rho(x) = \rho(x) - \rho_0$, where $\rho_0 \equiv \langle \rho(x) \rangle$ is the average density. The structure factor $S(q)$ in terms of the Fourier transform of $\delta \rho(x)$ is then
\begin{eqnarray} \label{eq:Sq_def_rho}
S(q) = \frac{1}{N} \left \langle |\delta \rho(q)|^2 \right \rangle,
\end{eqnarray}
where the brackets denote thermal averaging and $\delta \rho(q) = \rho(q) - \langle \rho(q) \rangle$ is the deviation of the Fourier-transformed density from its average value $\langle \rho(q) \rangle$. 

In the long wavelength limit $q \rightarrow 0$, we can directly calculate the structure factor $S(q)$ using Eq.~(\ref{eq:Sq_def_rho}). Mass conservation in one dimension requires the following relation between the displacement field and density fluctuations
\begin{eqnarray} \label{eq:mass_cons}
\delta \rho(x) = \rho_0 \partial_x u(x),
\end{eqnarray}
where $\rho_0 = \langle \rho(x) \rangle$ is the average density. 
In Fourier space, Eq.~(\ref{eq:mass_cons}) becomes
\begin{eqnarray}
\delta \rho(q) = \rho_0 i q u (q). 
\end{eqnarray}
From Eq.~(\ref{eq:deltaE_final}), the fluctuation energy can be written in terms of the density fluctuations as 
\begin{equation}
\Delta E = \int \frac{dq}{2 \pi} \frac{1}{2}  W(q) |\delta \rho(q)|^2 = \frac{1}{N}\sum_q \frac{1}{2} \frac{W(q)}{D} |\delta \rho(q)|^2, 
\end{equation}
where $W(q) =  B(q)/\rho_0^2$. With the help of Eq.~(\ref{eq:sum_j}), we obtain 
\begin{eqnarray}
W(q) =\frac{ 2 \pi A}{\rho_0^2  D^2} \frac{1}{q} = 2 \pi A \frac{1}{q},
\end{eqnarray}
to lowest order in $q$, where we have set $\rho_0 = N/L= D^{-1}$.  
Thus, the structure factor for small $q$, following Eq.~(\ref{eq:Sq_def_rho}), vanishes linearly in momentum
\begin{eqnarray} \label{eq:Sq_0}
\lim_{q \rightarrow 0} S(q) \approx \frac{k_B T}{A} \frac{D}{2 \pi}|q| = \frac{8 \pi k_B T }{ Yb^2} |\bar q|,
\end{eqnarray}
where $\bar q \equiv \frac{q}{2 \pi/D}$ is a dimensionless wavevector and we have used $A = Yb^2/8 \pi$. 
The vanishing of the structure factor as $q \rightarrow 0$ indicates the absence of long wavelength modes due to the incompressibility of dislocations with identical Burgers vectors, similar to a Coulomb gas of like-signed charges.  

\subsubsection{Bragg peaks}

In the previous section, we obtained the behavior of the structure factor near $q \rightarrow 0$ by directly computing the density-density correlation. A direct approach is more challenging at finite $q$, say, near a reciprocal lattice vector $G_m = \frac{2 \pi}{D} m$. To probe the structure factor near the wavevectors $\{ G_m \}$, we approximate this quantity using the one-dimensional displacement correlation function $C(s) \equiv \left \langle \left|u_s- u_0\right|^2 \right \rangle $.

To express the structure factor in terms of $C(s)$, we use Eq.~(\ref{eq:rho}) to rewrite Eq.~(\ref{eq:Sq_def_rho}) in the thermodynamic limit as~\cite{pathria2011statistical}
\begin{eqnarray} \label{eq:Sq_def}
S(q)= \left \langle \sum_{n=-\infty}^{\infty} \mathrm{e}^{i q\left(x_{n}-x_{0}\right)} \right \rangle, 
\end{eqnarray}
where $x_n$, the position of the $n$-th dislocation, can be decomposed into the equilibrium position $R_n = nD$ and a displacement $u_n$ as  $x_n = R_n + u_n$. On defining $s \equiv n - t$ and setting  $k \equiv q - G_m$, we can approximate Eq.~(\ref{eq:Sq_def}) for $|k| \ll G_m$ as
\begin{eqnarray} \label{eq:Sq_k}
S(G_m + k ) &=&  \sum_{s = -\infty}^{\infty} e^{i k D s} e^{- \frac{1}{2} G_m^2  \left \langle |u_s - u_0|^2 \right \rangle },
\end{eqnarray}
where we have used $e^{i G_m s D} = 1$ and the properties of Gaussian thermal averages to evaluate $\langle \exp \left[ i G_m (u_s - u_0) \right] \rangle$. 

Since the long wavelength modes provide the dominant contribution to the displacement correlation function $C(s)$, we calculate $\langle |u_s - u_0|^2 \rangle$ using Eq.~(\ref{eq:sum_j}) for large $s$. The displacement-displacement correlation is then
\begin{eqnarray}
C(s) &=& \left \langle \left|u_s- u_0\right|^2 \right \rangle  \label{eq:Gs_def}\\
&=& 2 \int_{-\pi/D}^{\pi/D} \frac{dq}{2 \pi} \int_{-\pi/D}^{\pi/D} \frac{dq'}{2 \pi} (1-e^{i q D s})  \\
&& \quad \quad \quad \quad \quad \quad \quad \quad \quad \quad  \times \left \langle u (q) u(q')\right \rangle  \notag\\
&=&  4 \frac{D^2 k_B T}{2 A \pi}   \int_{ 0}^{\pi/D} \frac{dq}{2 \pi}  \left( \frac{1-\cos(q D s)}{|q|} \right),\label{eq:int_whole}
\end{eqnarray}
where $ \langle u(q) u(q') \rangle$ has been evaluated using properties of thermal Gaussian averages and Eq.~(\ref{eq:deltaE_final}). 
In the limit of large $s \rightarrow \infty$, Eq.~(\ref{eq:int_whole}) can be approximated using cosine integrals~\cite{olver2010nist},
\begin{eqnarray} \label{eq:dispdisp}
C(s) &=& 4 \frac{D^2 k_B T}{2 A \pi}  \frac{1}{2 \pi} \left( \gamma + \ln(\pi s) + O \left[ \frac{\cos(\pi s)}{s} \right] \right),
\end{eqnarray}
where $\gamma \approx 0.577 $ is the Euler?Mascheroni constant. 
Upon substituting Eq.~(\ref{eq:dispdisp}) into Eq.~(\ref{eq:Sq_k}), we obtain the singular behavior of the structure factor for $|k| \ll G_m$, i.e., close to a reciprocal lattice vector, as
\begin{eqnarray}
 S(G_m + k) 
    &=&  \sum_{s = -\infty}^{\infty} e^{i k D s} e^{ - \gamma \frac{m^2 2 k_B T}{A } }  (\pi s)^{- \frac{ 2 m^2 k_B T }{ A}} \label{eq:sq_int}\\
       &=& \frac{A_m (T)}{(Dk)^{1-\alpha_m(T)}},  \label{eq:sq_div}
\end{eqnarray}
where the exponent $1-\alpha_m(T)$ is a temperature-dependent susceptibility critical exponent, with
\begin{eqnarray} \label{eq:sq_exp}
\alpha_m(T) = m^2 \frac{16 \pi k_B T}{Yb^2},
\end{eqnarray}
and the amplitude in Eq.~(\ref{eq:sq_div}) is
\begin{eqnarray}
 A_m(T) = \left( e^\gamma \pi\right)^{ -\alpha_m(T)} \int_{-\infty}^{\infty} d\eta e^{i \eta}  \eta^{- \alpha_m(T)} . 
\end{eqnarray}
It follows from Eq.~(\ref{eq:sq_div}) and (\ref{eq:sq_exp}) that the structure factor near the $m$-th reciprocal lattice vector $G_m$ has a singular contribution that scales according to
\begin{eqnarray} \label{eq:Sq_final}
\lim_{q \rightarrow G_m} S (q)  &\sim&\frac{1}{\left |q-G_m\right|^{1-\alpha_m(T)}}. 
\end{eqnarray}

We see from Eq.~(\ref{eq:Sq_final}) that at temperatures low enough such that ${1 - \alpha_m(T) \geq 0}$, the structure factor diverges as the momentum $q$ approaches the $m$-th reciprocal lattice vector (i.e. as $|k| =|q - G_m| \rightarrow 0$).  Thus, if we start from zero temperature (and neglect for now the pinning effect of the Peierls potential), there is an infinite set of diverging Bragg peaks, one located at every reciprocal vector $G_m$. When the pileup is in this floating-defect solid phase, these Bragg peaks decay algebraically as a function of $q = G_m - k$, with exponent $1 - \alpha_m(T) $ such that the higher order Bragg peaks are less singular than the more prominent ones closer to the origin in momentum space. The temperature-dependent exponents characterizing the divergence of the Bragg peaks in this quasi-long range ordered phase of the dislocation pileups are reminiscent of the Bragg peaks below the melting temperature of 2d crystals~\cite{nelson1979dislocation,halperin1978theory,kosterlitz1973ordering}. 
As the temperature increases, divergences in the highest order Bragg peaks vanish one by one upon surpassing the transition temperature $ \{ T_c^{(m)} \}$, where
\begin{eqnarray} \label{eq:Tcm}
k_B T^{(m)}_c = \frac{1}{m^2} \frac{Y b^2}{16 \pi}.
\end{eqnarray}
The last Bragg peak to disappear is the first-order Bragg peak at $G_1 = \frac{2 \pi}{D}$ closest to the origin in momentum space. Interestingly, the temperature at which this last remaining Bragg peak vanishes $k_B T_c^{(1)}$ coincides with the dislocation-unbinding temperature of the 2d host crystal, up to renormalizations discussed in Sec.~\ref{sec:gr}. Note that the spacing $D$ between dislocations in the pileup drops out in Eq.~(\ref{eq:sq_exp}) for the exponents $\{\alpha_m(T)\}$ and in Eq.~(\ref{eq:Tcm}) for the transition temperatures $\{T_c^{(m)} \}$. This independence of $D$ arises because the $D$-dependence of the interaction strength in Fourier space $B(q) \sim 1/D^2$ cancels against the $D$-dependence of the reciprocal lattice vectors $\{G_m \} = \{ m 2 \pi /D \}$. Similar results have been found for low angle grain boundaries in two dimensions~\cite{zhang2020LAGB}.

Upon combining our results for $q \approx 0$ and $q \approx G_m$ for $m = \pm 1, \pm 2, \cdots$, we expect the following form for the structure factor for $q > 0$, 
\begin{eqnarray} \label{eq:Sq_SC_full}
S(q) \approx S_0(q)  + \sum_{m=1}^\infty S_m(q),
\end{eqnarray}
where $S_0(q)$ is the dominant term in the limit $q \rightarrow 0$, and $S_m(q)$ is the dominant term in the limit $q \rightarrow G_m$, i.e.
\begin{eqnarray}
\lim_{q \rightarrow 0} S(q) = S_0(q), \quad \lim_{q \rightarrow G_m} S(q) = S_m(q). 
\end{eqnarray}Upon requiring $S(q)$ to be consistent with Eq.~(\ref{eq:Sq_0}), we immediately see that
\begin{eqnarray} \label{eq:S0}
S_0(q) = \frac{\alpha_1(T)}{2} |\bar q|,
\end{eqnarray}
while according to Eq.~(\ref{eq:Sq_final}), we expect
\begin{eqnarray} 
\lim_{q \rightarrow G_m} S_m(q) &\sim&\frac{1}{\left |q-G_m\right|^{1-\alpha_m(T)}}. 
\end{eqnarray}
We note that there are multiple ways to write $S_m(q)$ that would encompass the limiting behavior at $q \rightarrow G_m$.
To compare our theory with numerically extracted structure factors from random matrix simulations later in this section, we focus on the momentum range below the second reciprocal lattice vector $0<q < G_2$ and the temperature range $T_c^{(2)} >T \geq T_c^{(1)} $, where only the first order Bragg peak is divergent. In this regime, we will neglect contributions from higher order $m > 1$ Bragg peaks and decompose the structure factor $S(q)$ as
\begin{eqnarray} \label{eq:Sq_exact}
S(q) = S_0(q) + S_1(q). 
\end{eqnarray}
In the next section, with the help of random matrix theory, we write down an ansatz for $S(q)$ in this regime that combines the two scalings embodied in Eq.~(\ref{eq:Sq_0}) and Eq.~(\ref{eq:Sq_final}) with no fitting parameters. 

\begin{figure}[htb]
\centering
\includegraphics[width=1\columnwidth]{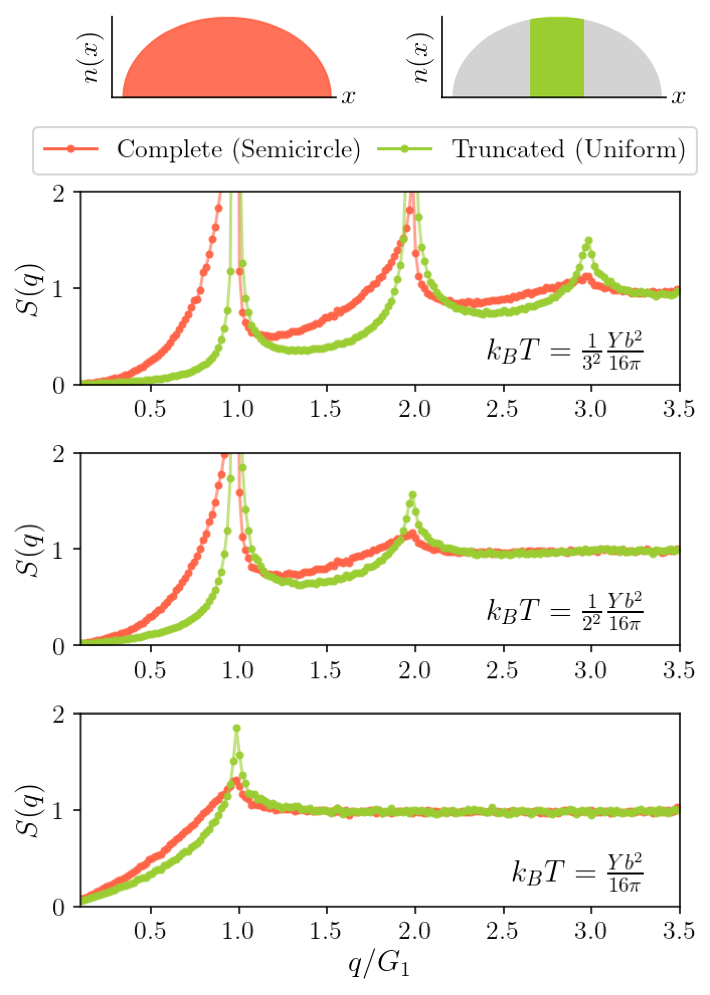}
\caption{\setstretch{1.} Structure factor $S(q)$ for the semicircular pileup (red) and approximately uniform pileup (semicircle pileup truncated to include only those dislocations within 25$\%$ of the center) (green) averaged over 500 realizations of rank $N=5000$ random matrices. The 1st, 2nd, and 3rd Bragg peaks form at $\beta = 4$, 16, 36, where $\beta$ is the dimensionless random matrix inverse temperature parameter $\beta = \frac{Yb^2}{4 \pi k_B T}$ in Eq.~(\ref{eq:bH_RMT}), as predicted by Eq.~(\ref{eq:Sq_final}). The wavevectors $q$ on the $x$-axes are scaled by the first reciprocal lattice vector $G_1 = 2 \pi/D$ where $D$ is the eigenvalue spacing at the center of the semicircular lattice. Straight lines connecting the dots are there to guide the eye. \label{fig:Sq_RMT}}
\end{figure}

\subsubsection{Connection to random matrix theory}

As discussed in the introduction, the dimensionless inverse temperature parameter $\beta$ in the $\beta$-Gaussian random matrix ensemble is given by $\beta = Yb^2/4 \pi k_B T$ (see also Eq.~(\ref{eq:bH_RMT})).  The temperature dependent critical exponent from Eq.~(\ref{eq:sq_exp}) can then be written in terms of this random matrix parameter $\beta$ as
\begin{eqnarray} \label{eq:alpha_beta}
\alpha_m(T) = \frac{4 m^2}{\beta}. 
\end{eqnarray}
The structure factor in the $q \rightarrow 0$ and $q \rightarrow G_m$ limits in Eqs.~(\ref{eq:Sq_0}) and (\ref{eq:Sq_final}) can also be expressed in terms of $\beta$ as
\begin{eqnarray}
\lim_{q \rightarrow 0} S(q) &= & S_0(q) = \frac{2}{\beta} \frac{q}{G_1}  \label {eq:S0_beta}\\ 
\lim_{q \rightarrow G_m} S(q) &\sim& \frac{1}{\left | G_m - q \right|^{1-\frac{4m^2}{\beta}}}. 
\end{eqnarray}
Note that the $m$-th order Bragg peak, centered at $G_m = m\frac{2 \pi}{D}$ for a uniform lattice, disappears when $\beta < \beta_c^{(m)}$, with
\begin{eqnarray} \label{eq:TcM}
\beta_c^{(m)} = 4 m^2. 
\end{eqnarray}

Conveniently, the exact form of the structure factor at three specific temperatures $\beta = 1, 2,4$ for the uniform lattice (derived by studying the semicircular Wigner distribution of eigenvalues in the limit of infinite length) can be obtained from conventional random matrix theory using orthogonal polynomials~\cite{mehta2004random}. These results are summarized in Table~\ref{tab:Sq}, where $\bar q \equiv q/G_1$ such that the $m$-th Bragg peak is centered at $\bar q=m$. We can see from Table~\ref{tab:Sq} that the random matrix theory results at $\beta=(1, 2, 4)$ ($T = (\frac{1}{4}, \frac{1}{2}, 1 ) \times T_c^{(1)}$) are consistent with theoretical results in the $q \rightarrow 0$ limit (Eqs.~(\ref{eq:Sq_0}) and (\ref{eq:S0_beta})):
\begin{eqnarray} \label{eq:Sq0}
\lim_{\substack{ \\ q \rightarrow 0}} S(q) = \frac{2}{\beta} |\bar q|.
\end{eqnarray}

In particular, the results from random matrix theory at $\beta = 4$ motivate us to propose an exact asymptotic expression for the structure factor near the first Bragg peak. To determine the form for $S_1(q)$ in Eq.~(\ref{eq:Sq_exact}), we require that the following conditions are satisfied: (1) $S_1(q)$ is consistent with Eq.~(\ref{eq:Sq_final}) in the $ q \rightarrow G_1$ limit, (2) $S_1(q)$ is subdominant to $S_0(q)$ in the $q \rightarrow 0$ limit, and (3) $S(q) = S_0(q) + S_1(q)$ reduces to the exact result from random matrix theory at $\beta = 4$ ($\alpha_1(T)=1$). 
Based on these three conditions, we conjecture that the contribution due to the first Bragg peak $S_1(q)$ can be written as
\begin{eqnarray} \label{eq:S1}
S_1(q) &=&  \left| \frac{\bar q}{2} \right|^{\alpha_1(T)} \frac{\alpha_1(T)}{2(1-\alpha_1(T))} \left[ \frac{1}{ ( 1-\bar q)^{1 - \alpha_1(T)}} - 1 \right]. 
\end{eqnarray}
One can verify that Eq.~(\ref{eq:S1}) satisfies the three conditions listed above. First, Eq.~(\ref{eq:S1}) indeed diverges as the appropriate power law near the first Bragg peak $\bar q \rightarrow 1$. This is apparent for $\alpha_1 > 1$. In the limit of $\alpha_1 \rightarrow 1$, one can use the following identity
\begin{eqnarray}
\lim_{p \rightarrow 0} \frac{1}{p} \left( \frac{1}{|k|^p} - 1 \right) = -\ln |k|
\end{eqnarray}
to see that $S_1(q)$ diverges logarithmically as $q \rightarrow G_1$. 
Second, in the limit of $q \rightarrow 0$, $S_1(q)$ scales as
\begin{eqnarray}
\lim_{\substack{\alpha_1 \rightarrow 1, \\ q \rightarrow 0}} S_1(q) = O(|\bar q|^{1+\alpha_1(T)}),
\end{eqnarray}
which is subdominant to $S_0(q) \sim k_B T |\bar q|$ for $T>0$ (recall that $\bar q = q/(\frac{2 \pi}{D} )$). Finally, for $\alpha_1(T) = 1$ ($T = T_c^{(1)}$ and $\beta = 4$), Eq.~(\ref{eq:S1}) reduces to the following
\begin{eqnarray}
\lim_{\substack{\beta \rightarrow 4}} S_1(q) = \frac{|\bar q|}{4} \ln |1 - |\bar q||,
\end{eqnarray}
which matches the exact result from random matrix theory in row 3 of Table~\ref{tab:Sq}. 

Upon combining Eqs.~(\ref{eq:S0_beta}), (\ref{eq:S1}) and (\ref{eq:alpha_beta}), our conjectured form for the structure $S(q)$ in the temperature range $T_c^{(2)} >T \geq T_c^{(1)} $ ($4 \leq \beta < 16$) can be expressed in terms of $\beta$ as
\footnotesize
\begin{eqnarray} \label{eq:Sq_RMT}
S(q) = \frac{2}{\beta} \left [ |\bar q| + \left| \frac{\bar q}{2} \right|^{4/\beta} \frac{1}{1-(4/\beta)} \left( \frac{1}{ ( 1-\bar q)^{1 - (4/\beta)}} - 1 \right) \right ].
\end{eqnarray} \normalsize
As shown in the next section, this expression shows excellent agreement with results from random matrix simulations.  

\begin{table}[h!]
\centering
\begin{tabular}{c|c}
$\beta$ & $S(q) \equiv K(\bar q), \quad \bar q \equiv \frac{q}{G_1}$  \\
\hline
1       & $\begin{array}{ll}{2|\bar q|-|\bar q| \ln (1+2|\bar q|),} & {|\bar q| \leq 1} \\ {2-|\bar q| \ln \left(\frac{2|\bar q|+1}{2|q|-1}\right),} & {|\bar q| \geq 1}\end{array}$ \\
\hline
2       & $\begin{array}{ll}{|\bar q|,} & {|\bar q| \leq 1} \\ {1,} & {|\bar q| \geq 1}\end{array}$\\
\hline
4       & $\begin{array}{ll}{\frac{1}{2}|\bar q|-\frac{1}{4}|\bar q| \ln |1-|\bar q||,} & {|\bar q|\leq 2} \\
{1,} & {|\bar q| \geq 2} \end{array}$                                                         
\end{tabular}
\caption{Exact expressions for $S(q) \equiv  K\left (\bar q\right)$, where $D$ is the dislocation spacing and $\bar q = q / G_1 = q D/2 \pi$, derived from random matrix theory via orthogonal polynomials for the special values of the dimensionless random matrix inverse temperature parameter $\beta = 1, 2, 4$~\cite{mehta2004random}.}
\label{tab:Sq}
\end{table}

\subsection{Random matrix simulations \label{sec:RMS_uni}}

In this section, we compute the structure factors numerically for a system of $N$ dislocations using~\cite{pathria2011statistical} 
\begin{eqnarray}
S(q) = \frac{1}{N} \left \langle \sum_{n,m} e^{-iq(x_n - x_m)} \right \rangle,
\end{eqnarray}
with the dislocation positions given by the eigenvalues of random matrix simulations, and show that they agree with our theory from the previous section. 

Figure \ref{fig:Sq_RMT} shows the structure function $S(q)$ for the semicircle lattice of eigenvalues (red), averaged over 500 realizations of rank $N=5000$ random matrices $H_\beta$ from Eq.~(\ref{eq:Hb}), with $\beta = 4, 16, 36$, the random matrix inverse temperatures $\beta$ at which the first, second, and third Bragg peaks start exhibiting a power law divergence for a uniform pileup according to Eq.~(\ref{eq:Sq_final}) and Eq.~(\ref{eq:TcM}). Note from e.g. Eqs.~(\ref{eq:S1}) and (\ref{eq:Sq_RMT}) that we expect power law divergences at the Bragg peaks precisely at these special values of $\beta$. Note also that the Bragg peaks in red are asymmetric, with their more pronounced wings extending towards $q=0$. This feature arises from the longer lattice spacings near the edges of the semicircle lattice, where the dislocations corresponding to the eigenvalues are more dilute. 

To better compare the random matrix eigenvalues to our theory of the \textit{uniform} pileup, we also show results from truncating the semicircle lattice such that only the $N/4$ dislocations closest to the center of the lattice are retained (see Fig.~\ref{fig:Sq_RMT}). In this interval, the dislocation/eigenvalue spacings are approximately constant and the lattice is approximately homogeneous. The resulting structure factors $S(q)$ are shown in Fig.~\ref{fig:Sq_RMT} in green. After the truncation, which removes the dislocations with longer lattice spacings, the asymmetric wings on the inner edges of the Bragg peaks vanish, and the Bragg peaks are more centered at $q/G_1 = 1, 2, 3$. 

\begin{figure*}[tb]
\centering
\includegraphics[width=0.95\textwidth]{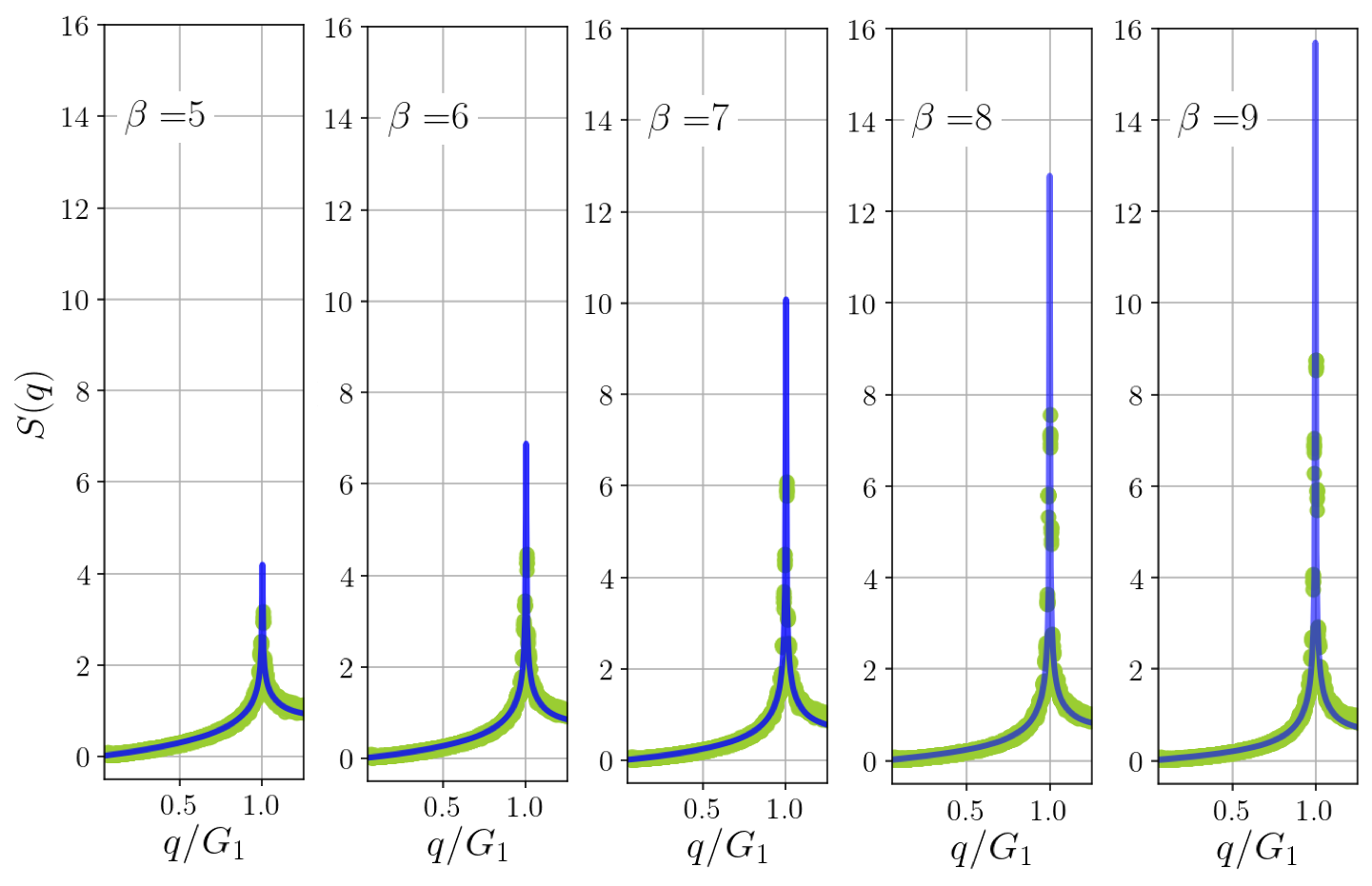}%
\caption{\setstretch{1.} Structure factor $S(q)$ of the truncated semicircle lattice (i.e. approximately uniform lattice, see schematic on top right of Fig.~\ref{fig:Sq_RMT})b) near the first reciprocal lattice vector at different temperatures in range $T^{(2)}_c > T \geq T^{(1)}_c$ ($16 > \beta \geq 4$). Green dots show the results from random matrix simulations while blue lines are from Eq.~(\ref{eq:Sq_RMT}); simulations and theory show good agreement. \label{fig:Sq_power_trunc}}
\end{figure*}

We can now compare the numerical results with the our theory summarized in Eq.~(\ref{eq:Sq_RMT}). As shown in Fig.~\ref{fig:Sq_power_trunc}, Eq.~(\ref{eq:Sq_RMT}), an exact expression with no fitting parameter, agrees well with the structure factors extracted from random matrix simulations for $\beta > \beta_c^{(1)}$ near the first reciprocal lattice vector. 

We find that the heights of the Bragg peaks extracted from the random matrix eigenvalues are approximately constant as a function of system size $N$, provided that we average over a large number of realizations. Although for uniform dislocation lattices, one might expect the height of the Bragg peak to scale as $S(G_1) \sim N^{\alpha-1}$ by setting $k\approx \pi/N$ in Eq.~(\ref{eq:sq_int}), this result is not confirmed by random matrix simulations. This discrepancy may arise because the $\sim N^{\alpha-1}$ scaling requires a uniform lattice constant $D(x) = D$ over the entire sample size, leading to a precisely defined reciprocal lattice vector $G_1(x) = G_1 = 2 \pi/D$ for all $x$. For our general $\beta$-Gaussian random matrix simulations, although the truncated semicircle lattice is sufficiently uniform such that $S(q)$ has the correct power law divergence behavior near the Bragg peaks, the equilibrium lattice spacing still has nonzero variation given by a slowly varying function $D(x)$. Although small, this variation is enough to smear out the very tip of the Bragg peak, which is sensitive to a wide range of dislocation spacing $D(x)$. 

\subsection{Structure factor for inhomogeneous pileups \label{sec:Sq_inhomo}}
\begin{figure*}[t]
\centering
\includegraphics[width=0.95\textwidth]{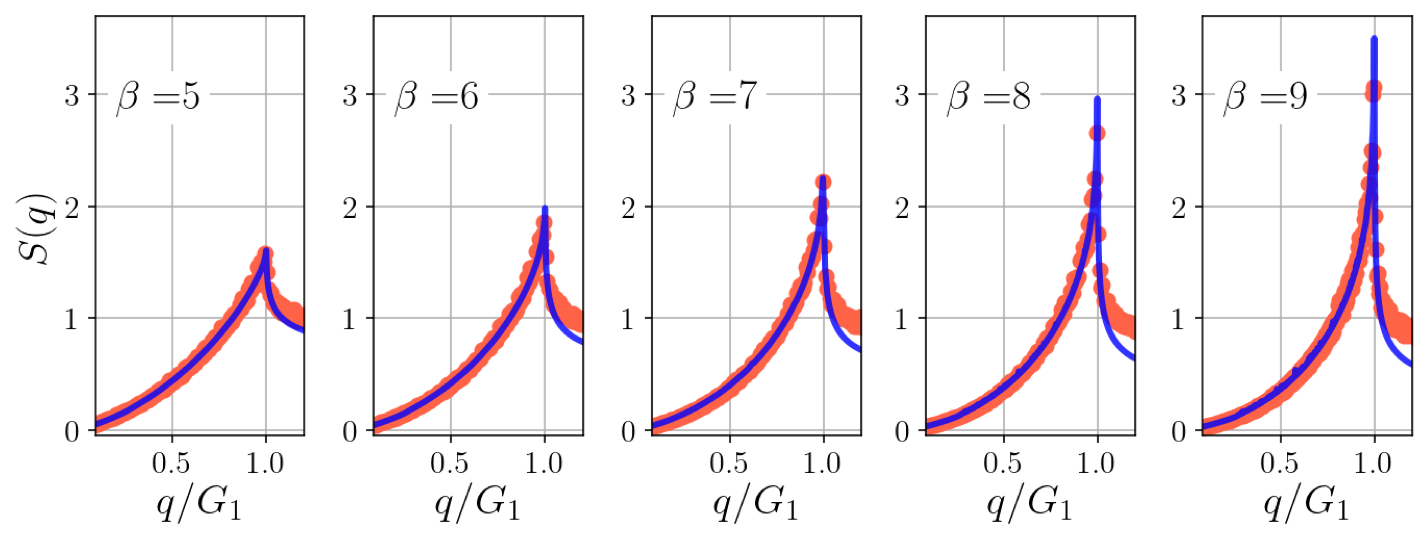}%
\caption{\setstretch{1.} Structure factor $S(q)$ of the full semicircular pileup at different temperatures in range $T^{(2)}_c > T \geq T^{(1)}_c$ ($16 > \beta \geq 4$). Red dots show the results from random matrix simulations while blue lines are Eq.~(\ref{eq:Sq_SC_test}); simulations and theory show good agreement. Deviations above the first Bragg peak $q > G_1$ likely come from the smeared out contribution of the $S_2(q)$ term, neglected in Eq.~(\ref{eq:Sq_exact}), which contributes a noise floor of $S(q \rightarrow \infty) = 1$ in the absence of higher order Bragg peaks. Here, $G_1$ is defined as $G_1 \equiv 2 \pi/D(x=0)$, where $D(x=0)$ is the average dislocation spacing at the center of the semicircular pileup.  \label{fig:Sq_SC}}
\end{figure*}
For an inhomogeneous pileup with spatially varying lattice constant $D(x)$, we can calculate the overall structure factor via a direct one-dimensional ``powder average.''  Upon defining $K(\frac{q}{G_1}) \equiv S(q)$, the structure factor for an inhomogeneous pileup with a variable average dislocation spacing $D(x)$ is then approximately
\begin{eqnarray} \label{eq:Sq_powder_cont}
S(q)  &=& \frac{1}{N}\int dx  \bar n(x) K\left (\frac{q}{2 \pi / \bar D(x)} \right),
\end{eqnarray}
where $\bar n(x) \equiv 1/\bar D(x)$ is a coarse-grained density. If the form of a smooth, continuous, density $n(x)$ is known, then one can simply set $\bar D^{-1}(x) =\bar n(x) = n(x)$. In practice, given a set of discrete dislocation positions $\{ x_i \}$, the structure factor is given by
\begin{eqnarray} \label{eq:Sq_SC_test}
S(q) =\frac{1}{N}\sum_{i=1}^N K\left (\frac{q}{2 \pi / \bar D(x_i)} \right),
\end{eqnarray}
where the coarse-grained lattice constant can be estimated using a sliding window to average over, say, 5 neighboring lattice sites according to $\bar D(x_i) = \frac{1}{5}\sum_{|j-i|\leq 2} D(x_j)$, determining the local reciprocal lattice vector as $G_1(x_i) = 2 \pi/D(x_i)$. 

To test our theory when averaged over inhomogeneous dislocation spacings, we again focus on the $0<q < G_2$ momentum range, and the temperature range $T^{(2)}_c > T \geq T^{(1)}_c$ (random matrix parameter range $16 > \beta \geq 4$) where the first order Bragg peak dominates. The structure factor under these conditions is then given by the inhomogeneous lattice average in Eq.~(\ref{eq:Sq_SC_test}) with
\begin{eqnarray} 
K &&\left( \frac{q}{2 \pi/\bar D(x) } \right) = \frac{\alpha_1(T)}{2}  \Bigg \{ \left |\frac{q}{2 \pi/\bar D(x) } \right| \\
&& + \left| \frac{q}{2 (2 \pi/\bar D(x)) } \right|^{\alpha_1(T)} \frac{1}{(1-\alpha_1(T))} \notag  \\
&& \times \Bigg[ \left ( 1-\frac{q}{2 \pi/\bar D(x) } \right)^{-(1 - \alpha_1(T))} - 1 \Bigg] \Bigg \}. \notag 
\end{eqnarray}

As seen in Fig.~\ref{fig:Sq_SC}, Eq.~(\ref{eq:Sq_SC_test}) shows good agreement with the structure factor extracted from random matrix simulations of the inhomogeneous semicircular pileup. The deviation past the first Bragg peak $q > G_1$ likely comes from the smeared out contribution of the $S_2(q)$ term neglected in Eq.~(\ref{eq:Sq_exact}), which contributes a noise floor of $S(q \rightarrow \infty) = 1$ in the absence of higher order Bragg peak divergences. Note that for a general inhomogeneous pileup, the reciprocal lattice vector $G_m$ is not well defined since the average dislocation spacing $D(x)$ varies in space. In Fig.~\ref{fig:Sq_SC}, $G_1$ is defined as $G_1 \equiv 2 \pi/D(x=0)$, where $D(x=0)$ is the average dislocation spacing at the center of the semicircular pileup, also the densest region in the semicircular pileup. 

We note that Bragg peaks of an inhomogeneous pileup can become difficult to detect when the lattice spacing $D(x)$ does not have a finite lower bound. Since $n(x) \sim D(x)^{-1}$, structure factors like that in Eq.~(\ref{eq:Sq_powder_cont}) can be dominated by signals from the portions of the lattice with very small lattice spacings, corresponding to large reciprocal lattice vectors. If $D(x)$ goes to 0 at some location in the pileup, the Bragg peaks in the structure factor run away to arbitrarily large values $\frac{2 \pi}{D(x)}$. We find that this anomalous behavior arises for the single pileup and the double pileup shown in Table~\ref{tab:disloc}, where the lattice constant $D(x)$ goes to zero as the density $n(x)$ diverges towards the pileup edges. Nevertheless, one could in principle detect signatures of algebraic long-range order in these pileups by measuring the structure factor of locally crystalline segments small enough such that the lattice constant is approximately uniform within the segment. We can also probe quasi-long range order in inhomogeneous lattices by studying the local radial distribution function $g(r)$, as shown in the next section. 

\subsection{Radial distribution function \label{sec:gr}}
In this section, we examine the dislocation ordering in pileups using a quantity complementary to the structure factor, the radial distribution function, also called the pair correlation function or the two-point correlation function. For a particular realization of dislocations extracted from, say, a random matrix ensemble, the radial distribution function $g(r)$ determines the probability of finding a second dislocation a distance $r$ away from some first existing dislocation. Scaling arguments based on the structure factor $S(q)$ derived in the previous section and results from random matrix theory allow us to identify oscillations in $g(r)$ that decay algebraically as a function of inter-dislocation distance $r$, the signature of quasi-long range order in real space~\cite{nelson2002defects}. These oscillations are controlled by the same exponents $\{\alpha_m \}$ that determines the divergences in the Bragg peaks of the structure factors discussed in the previous subsections. 

\begin{table}[h!]
\centering
\begin{tabular}{c|c}
$\beta$ & $ g(r) = h(\bar r), \quad \bar r = \frac{r}{D} $\\
\hline
1       & \begin{tabular}[c]{@{}c@{}}
$1-\left(\int_{\bar r}^{\infty} s(\bar t) d \bar t\right)\left(\frac{d}{d r} s(\bar r)\right)+(s(\bar r))^{2}, \quad s(\bar r) = \frac{\sin \pi \bar r}{\pi \bar r}$\\ 
Large $\bar r$: $1-\frac{1}{\pi^{2} \bar r^{2}}+ \frac{3}{2 \pi^4 \bar r^4} + \frac{\cos 2 \pi \bar r}{\pi^{4} \bar r^{4}}+\cdots$ \end{tabular}   \\
\hline
2       & \begin{tabular}[c]{@{}c@{}}
$1- s(\bar r)^{2}$ \\ 
Large $\bar r$: $1-\frac{1}{2 \pi^{2} \bar r^{2}}+ \frac{\cos 2 \pi \bar r}{2 \pi^{2} \bar r^{2}} + \cdots $\end{tabular}   \\
\hline
4       & \begin{tabular}[c]{@{}c@{}}

$1-s(2 \bar r)^{2}+\frac{d}{d \bar r}s(2 \bar r)\cdot \int_{0}^{\bar r}s(2 \bar t) d \bar t $\\ 
Large $\bar r$: $1+\frac{\pi}{2} \frac{\cos (2 \pi \bar r)}{2 \pi \bar r}+\cdots$\end{tabular}
\end{tabular}
\caption{Exact expressions for $g(r) = h(\bar r)$ from random matrix theory derived via orthogonal polynomials, where $\bar r=\frac{r}{D}$ scales the inter-eigenvalue distance $r=\left|x_{1}-x_{2}\right|$ by the mean eigenvalue spacing near the center of the semicircle lattice $D$, and $s(\bar r) \equiv \frac{\sin \pi \bar r}{\pi \bar r}$~\cite{mehta2004random}.}
\label{tab:gr}
\end{table}

\begin{figure}[htb]
\centering
\includegraphics[width=0.95\columnwidth]{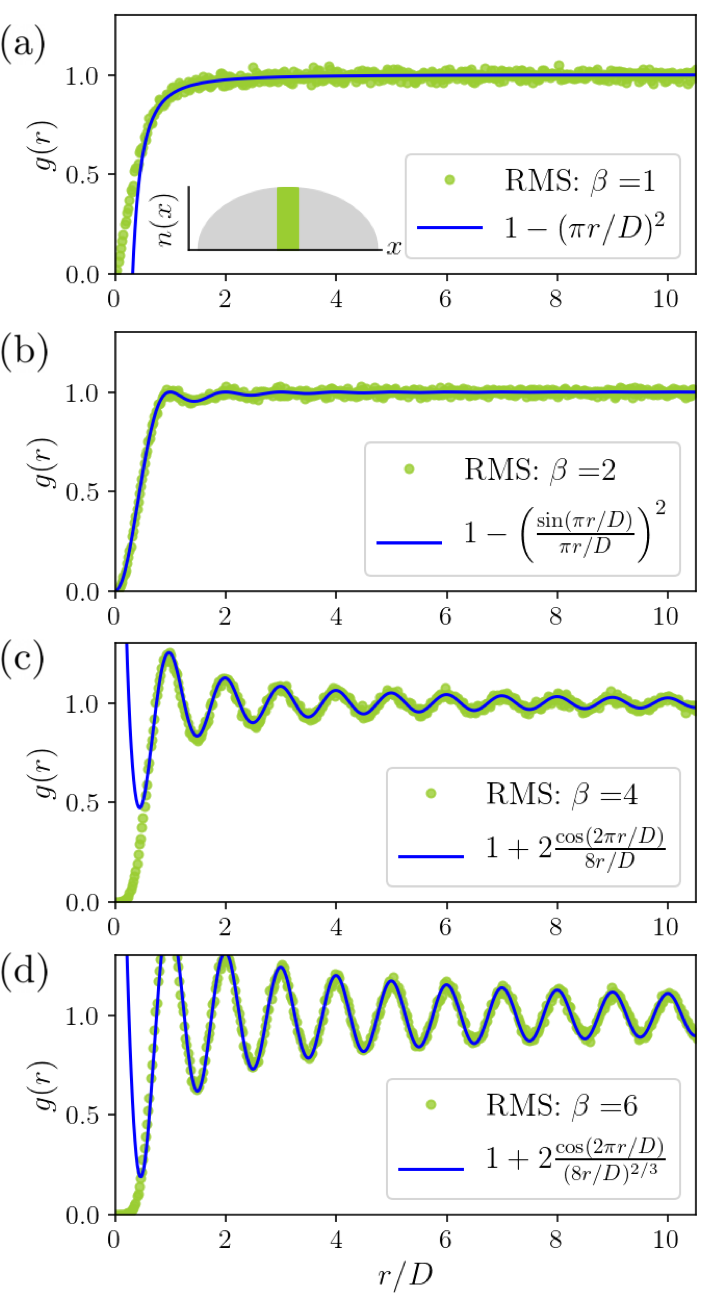}
\caption{\setstretch{1.} Radial distribution function $g(r)$ at $\beta = 1, 2, 4, 6$ for uniform pileups. This quantity always approaches unity for large $r$. Green dots are data from random matrix simulations (RMS) of the truncated semicircle pileup (i.e. an approximately uniform pileup, see see inset of (a)) averaged over 500 realizations of rank $N=5000$ random matrices. Large $\frac{r}{D}$ behaviors from random matrix theory at $\beta = 1,2,4$ (Table~\ref{tab:gr}) are plotted in blue in (a)-(c) ($D = \frac{\pi}{2 N}$ is known from random matrix theory to be the mean lattice spacing at the middle of the semicircle lattice spanning $ (-1, 1)$). Large $\frac{r}{D}$ behavior according to Eq.~(\ref{eq:gr_exact}) at $\beta = 6$ is plotted in blue in (d). \label{fig:gr_RMT}}
\end{figure}

The two-point correlation functions from random matrix theory for the uniform lattice (derived by studying the semicircular eigenvalue distribution in the limit of infinite length) at the special dimensionless inverse temperature parameter $\beta = \frac{Yb^2}{4 \pi k_B T}= 1,2,4$, accessible via conventional random matrix theory, are shown in Table~\ref{tab:gr}, along with their behavior when the separation distance $r$ is large relative to the lattice spacing $D$~\cite{mehta2004random}. Upon denoting $\bar r \equiv \frac{r}{D}$ as the separation distance scaled by the dislocation spacing $D$ at the center of the band and letting $h(\bar r) \equiv g(r)$, we can see from Table~\ref{tab:gr} that the leading terms at large $\bar r$ for $\beta = 1, 2$ are not oscillatory ($\sim \frac{1}{\bar r^2}$), while the leading term at large $\bar r$ for $\beta = 4$ is oscillatory with a periodicity corresponding to the first reciprocal lattice vector ($\sim \frac{ \cos(2 \pi \bar r)}{\bar r}$). Thus, although there are algebraically-decaying oscillatory terms $\sim \cos 2 \pi \bar r$ in the correlation function at $\beta<4$, they are sub-dominant to a \textit{non-oscillating} algebraically decaying term, consistent with no appreciable translational order at high temperatures. When $\beta =1$, the oscillatory term is suppressed so much that it is invisible to the eye (Fig.~\ref{fig:gr_RMT}a), while for $\beta = 4$, the oscillations $\sim \cos 2 \pi \bar r$ are visibly modulated by an algebraically decaying envelope symmetric about $g(r) = 1$. 

We can compare the correlation function $g(r)$ corresponding to our theory of dislocation pileups as a function of temperature $T$, by Fourier transforming the structure factor $S(q)$:
\begin{eqnarray}
g(r)-1 = \frac{1}{\rho_0} \int dq \left[S(q) -\delta(q) - 1\right]e^{i q r},
\end{eqnarray}
where $\rho_0$ is the average dislocation density. 
For $T_c^{(1)} \geq T > T_c^{(2)}$, we expect that the contribution from the first Bragg peak dominates for small $k \equiv q - G_1 $, and we can obtain the shape of the envelope modulating the oscillations in $g(r)$ as
\begin{eqnarray}
g(r \neq 0) - 1 &=& \frac{1}{\rho} \int dq ~S(q) e^{i q r}, \\
&\sim& e^{-i G_1 r}\int dk ~|k|^{-1+\alpha_1(T)}e^{i k r} \\
    &\sim& r^{-\alpha_1(T)} \cos (G_1 r). \label{eq:gr_rough_scale}
\end{eqnarray}
This scaling gives the power law decay of correlations in real space, written in usual critical phenomena conventions as $g(r) \sim 1/r^{d-2+\eta}$~\cite{stanley1987introduction}, where $d=1$, $g(r) \sim 1/r^{\alpha_1(T)}$, with $\alpha_1(T)$ given by Eqs.~(\ref{eq:sq_exp}) and (\ref{eq:alpha_beta}). 

Equation (\ref{eq:gr_rough_scale}) also displays oscillatory behavior given by $G_1 = 2 \pi /D$ on the scale of the dislocation spacing $D$, so the radial distribution function behaves at large $r$ according to
\begin{eqnarray} \label{eq:gr_scale}
\lim_{r \rightarrow \infty} (g(r) -1) \sim \frac{\cos (2 \pi r/D)}{(r/D)^{\alpha_1(T)}}. 
\end{eqnarray}Upon utilizing the exact result for $\beta = 4$ ($\alpha_1(T) = 1$) in row 3 of Table~\ref{tab:gr}, we determine the coefficients of Eq.~(\ref{eq:gr_scale}) up to a single fitting parameter $c$:
\begin{eqnarray} \label{eq:gr_fit}
\lim_{r \rightarrow \infty} g(r) = 1 + \frac{c}{4} \frac{\cos\left (2 \pi r/D\right )}{\left( c r/D\right)^{\alpha_1(T)}}. 
\end{eqnarray}
Upon examining the entire temperature range  $T_c^{(1)} \geq T > T_c^{(2)}$ ($16 > \beta \geq 4$), we find that $c \approx 8.0$, so the radial distribution function at large $r$ takes the form
\begin{eqnarray} \label{eq:gr_exact}
\lim_{r \rightarrow \infty} g(r) = 1 + 2 \frac{\cos\left (2 \pi r/D\right )}{\left( 8 r/D\right)^{\alpha_1(T)}}. 
\end{eqnarray}
Figures \ref{fig:gr_RMT}c-d and \ref{fig:gr_loc}a show excellent agreement between Eq.~(\ref{eq:gr_exact}) and the radial distribution functions extracted from random matrix simulations at $\beta = 4, 6, 8$, corresponding to low temperatures.

We also show the radial distribution functions for random matrix parameters $\beta = 1$ and $\beta = 2$ in Figs.~\ref{fig:gr_RMT}a-b for completeness. However, random matrix simulations of the 1d Coulomb gas at $\beta < 4$ correspond to dislocation pileup lattices at temperatures $ T > T_c^{(1)}$. The latter situation is likely inaccessible on flat two-dimensional host crystals, which are unstable to a dislocation unbinding mechanism at about the same temperature as the pileup they host $T_m \approx T_c^{(1)}$~\cite{nelson1979dislocation,nelson2002defects,halperin1978theory,kosterlitz1973ordering}. We expect that the Young's modulus of the 2d host crystal is renormalized by the dislocation pairs that unbind near melting: $ Y \rightarrow Y_R(l) $, where $l = \ln (L_\text{host}/a)$ and $L_\text{host}$ is the host crystal size while $a$ is the host lattice constant. As a result, the melting temperature $T_m$ of the 2d host crystal is shifted as
\begin{eqnarray}
T_m = \frac{Y_R(l) b^2}{16 \pi}. 
\end{eqnarray}
The unbinding dislocation pairs in the host crystal will also renormalize the interactions between the dislocations in a pileup. Since the physics of pileup melting is dominated by interactions at long wavelengths $B(q) \sim |q|^{-1}$ (see Eq.~(\ref{eq:deltaE_final})), their melting temperature will also be shifted by a partially renormalized Young's modulus
\begin{eqnarray}
T_c^{(1)} \approx \frac{Y_R(l') b^2}{16 \pi},
\end{eqnarray}
where $l' = \ln (L/a)$ and $L$ is the length of the pileup. It seems plausible that the renormalized Young's modulus $Y_R(l')$ that controls the physics of the pileup is the same as the renormalized Young's modulus $Y_R(l)$ that controls the 2d melting transition of the host crystal \textit{if} the pileup and the host crystal are comparable in size $L \sim R$. If the host crystal is appreciably larger, it may melt sooner than the highest dislocation pielupe transition, corresponding to the loss of a diverging Bragg peak at $q = G_1$. 
Curved 2d crystals and their pileups (see Fig.~\ref{fig:intro_schm}b) may exhibit different renormalized Young's moduli depending on their geometry. It is conceivable that the disappearance of the first Bragg peak of a pileup at $T>T_c^{(1)}$ might be observable in this case. 

\subsubsection{Local radial distribution functions for inhomogeneous pileups}

\begin{figure}[htb]
\centering
\includegraphics[width=0.95\columnwidth]{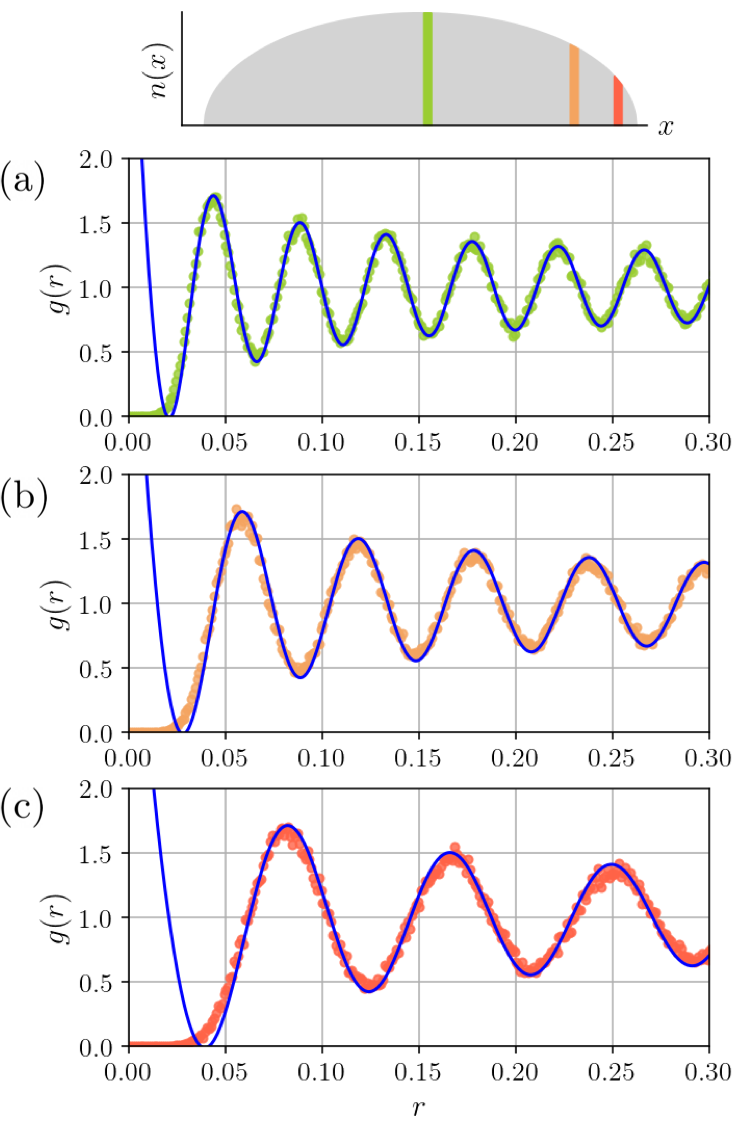}
\caption{\setstretch{1.} Local radial distribution functions $g_R(r)$ at $T = \frac{1}{2} T_c^{(1)}$ ($\beta = 8$) for inhomogeneous pileups at three different mean locations $R = \{ 0, 0.47, 0.85\}$ in the semicircular pileup of length $L = 2$ spanning $(-1,1)$, as indicated in the top graph. Dots are data averaged over 500 random matrix simulations using rank $N=5000$ matrices, colored according to the mean location $R$ indicated by the colored slices in the top schematic. Blue lines are the predictions of Eq.~(\ref{eq:gr_loc}) with the mean lattice spacing $\bar D(R)$ extracted from simulations by averaging the positions of dislocations (eigenvalues) within each region $R \pm 0.01$. Theory and simulations show good agreement. \label{fig:gr_loc}}
\end{figure}

It is known from random matrix theory that the two-point correlation function $g(r)$ is stable under translations over the eigenvalue spectrum provided that the distances $r$ are expressed in terms of the local mean eigenvalue spacing $D$~\cite{mehta2004random}. Thus, we can directly apply the above results for the radial distribution function of uniform pileups to the \textit{local} radial distribution function of inhomogeneous pileups. In particular, we expect that the local radial distribution function~\cite{zhang2019eigenvalue} of an inhomogeneous pileup $g_R( r)$, where $r = |x_1-x_2|$ is the inter-dislocation distance and $R = \frac{x_1 + x_2}{2} $ is the mean location of a pair of dislocations, may be written as
\begin{eqnarray} \label{eq:gr_loc}
g_R(r)=  h\left( \frac{r}{\bar D\left(R\right) } \right),
\end{eqnarray}
where $\bar D(R)$ is the mean lattice spacing at position $R$ as defined in Sec.~\ref{sec:Sq_inhomo}, and  $h(\bar r = \frac{r}{D}) \equiv g(r)$ as defined in Table~\ref{tab:gr} where $g(r)$ is the radial distribution function for a uniform dislocation lattice. As shown in Fig.~\ref{fig:gr_loc}, Eq.~(\ref{eq:gr_loc}) agrees well with random matrix simulations for the inhomogeneous semicircular pileup at $T = \frac{1}{2}T_c^{(1)}$ ($\beta = 8$). As we move towards the outer edge of the semicircular pileup, the dislocation spacing increases on average, but the shape of the local radial distribution function remains the same up to stretching along the $x$-axis. 

\section{Pinned-defect to floating-defect transition \label{sec:host}}
We have thus far treated the dislocations as interacting particles and the underlying host crystal as a continuous elastic medium. This is a valid approximation in the floating-defect phase, where the dislocations are unaffected by the lattice structure of the host crystal. In this section, we examine the effect of Peierls potential and identify a low-temperature transition from the floating-defect phase to a pinned-defect phase, where the physics is dominated by the underlying lattice potential of the host crystal and the one-dimensional pileup exhibits true long range translational order. As we will show, the structure factor of the pileup in the low-temperature pinned-defect phase exhibits delta function Bragg peaks at the reciprocal lattice vectors, in contrast to the power law Bragg peaks characterizing the floating defect phase. 

We first study pileups whose zero-temperature equilibrium density is accidentally commensurate with the host crystal (see illustration in Fig.~\ref{fig:dp}a). 
We show that a model of quantum Brownian motion model for a particle in a periodic potential in imaginary time maps directly onto the classical statistical mechanics of commensurate pileups. In particular, upon translating the renormalization group results from Refs.~\cite{fisher1985quantum} and \cite{schmid1983diffusion}, we can predict the transition temperature $T_P^0$ below which the floating-defect state becomes unstable to a pinned-defect state for accidentally commensurate pileups. This mapping onto a quantum problem shows that the long range interaction between the dislocations is crucial to the existence of the pinned defect phase; the repulsive logarithmic interaction potential is analogous to the friction force that prevents the quantum particle from tunneling to the other minima in the periodic potential. 
We then discuss \textit{incommensurate} pileups, reminiscent of the commensurate-incommensurate transition of adsorbed monolayers on a periodic substrate studied in Ref.~\cite{halperin1978theory} (see also Ref.~\cite{nelson2002defects}), where the transition temperature $T_P$ is depressed from $T_P^0$ by incommensurability. 

\subsection{Commensurate pileups and mapping to quantum Brownian motion}

\begin{table}[tb]
\centering
\begin{tabular}{c|c}
\hline \hline
quantum Brownian motion & dislocation pileup \\
\hline
action $ S$ & reduced Hamiltonian $H/k_B T$ \\
imaginary time $\tau$ & spatial coordinate $x$ \\
frequency $\omega$ &momentum $q$ \\
quantum particle position & dislocation displacement field\\ 
 $x(\tau)$ & $u(x)$\\
friction coefficient $\eta$ & $\frac{1}{k_B T}\frac{Y b^2}{4 \pi D^2}  $          \\
\hline \hline
\end{tabular}
\caption{Correspondence between quantum Brownian motion and the statistical mechanics of dislocation pileup.}
\label{tab:qbm}
\end{table}

The discrete nature of the atomic planes on either side of a dislocation produces Peierls potential, describing the preferred dislocation positions~\cite{hirth1983theory}. If the equilibrium positions of the dislocations are as to produce an integer number of extra atomic planes in between neighboring dislocations, then we say the pileup is ``accidentally commensurate'' with the host lattice (see Fig.~\ref{fig:dp}a). Upon incorporating Peierls potential into Eq.~(\ref{eq:deltaE_final}), the reduced Hamiltonian for an accidentally commensurate pileup becomes
\begin{eqnarray} \label{eq:H_P}
\frac{H}{k_B T} &=&  \frac{1}{k_B T}\frac{1}{2} \left( \frac{Y b^2}{4  D^2} \right) \int \frac{dq}{2 \pi} |q| |u(q)|^2 \\ 
&& - V_0 \Lambda \int dx \cos \left (\frac{ 2 \pi u(x) }{a} \right) \notag
\end{eqnarray}
where $a$ is the lattice constant of the host crystal, $V_0 = \frac{V_\text{Peierls}}{k_B T} \Lambda^{-1}$ where $V_\text{Peierls} $ scales with the energy difference between the highest energy location and lowest energy location of a dislocation in a unit cell of the host crystal and $\Lambda \sim a^{-1}$ is a short distance cutoff. (Higher order cosines in the Peierls potential, such as a coupling $\sim \cos (4 \pi u(x) / a)$, could be included, but these are less important than the terms we have kept.) The statistical mechanics associated with Eq.~(\ref{eq:H_P}) can be mapped onto the problem of a quantum particle experiencing friction in a periodic potential, also known as the quantum Brownian motion (QBM) model~\cite{fisher1985quantum,schmid1983diffusion}. The correspondence is detailed in Table~\ref{tab:qbm}. 

Upon treating $V_0$ as a perturbation to the floating defect Hamiltonian in Eq.~(\ref{eq:H_P}), the renormalization group recursion relation for $V_0$ reads~\cite{fisher1985quantum,schmid1983diffusion}
\begin{eqnarray} \label{eq:Va}
\frac{d V_{0}(l)}{d l}=\left(1-\frac{1}{\gamma}\right) V_{0}(l),
\end{eqnarray}
where $l$ determines the fraction of short wavelength degrees of freedom that have been integrated out in the coarse-graining procedure and
\begin{eqnarray}
\gamma = \frac{1}{k_B T}\frac{Y b^2 a^2}{8 \pi D^2}. 
\end{eqnarray}
The condition $\gamma = 1$ then gives us the temperature above which the Peierls potential strength $V_0$ becomes irrelevant to the long wavelength statistical mechanics,
\begin{eqnarray} \label{eq:TF0}
k_B T_P^0 = \frac{Yb^2}{16 \pi} \frac{2 a^2}{D^2}. 
\end{eqnarray}
Below the pinning temperature $T_P^0$, Peierls potential is relevant and $V_0$ iterates to $\infty$ at long wavelengths. In this pinned-defect phase, dislocation motion in the glide plane is frozen out and these defects are pinned in place. Above $T_P^0$, Peierls potential is irrelevant and $V_0$ iterates to $0$, leading to the floating-defect phase, where dislocations behave as logarithmically interacting Coulomb charges in 1d and exhibit the quasi-long range order described in the previous section. 
Recall that the first Bragg peak no longer diverges for $T>T_c^{(1)} = Yb^2/16 \pi$ from Eq.~(\ref{eq:Tcm}). Since dislocations in a pileup are typically separated by many host lattice spacings $(a/D)^2 \ll 1$, we expect
\begin{eqnarray}
T_P^0 \ll T_c^{(1)}. 
\end{eqnarray}
Thus, a large temperature range exists between $T_P^0$ and $T_c^{(1)}$ where the host lattice structure is irrelevant and the unmodulated Coulomb gas-like interactions between dislocations dominate the physics of pileups.

In the QBM model, $V_0 \rightarrow 0$ corresponds to the delocalized phase of the quantum particle, while $V_0 \rightarrow \infty$ corresponds to the localized phase, where the particle is arrested by friction and cannot tunnel through the walls of the periodic potential from one minimum to another. As summarized in Table~\ref{tab:qbm}, the friction term in the QBM problem maps onto the long range interaction term in the pileups Hamiltonian, so the repulsive logarithmic interaction potential is analogous to the friction force. Thus, the repulsive long range interaction between the dislocations cooperates with Peierls potential to keep the dislocations locked in place; long range interaction with other dislocations and Peierls potential are both necessary for the pinned defect phase to exist at finite temperatures. 

\subsection{Incommensurate pileups}

\begin{figure}[h]
\centering
\includegraphics[width=0.9\columnwidth]{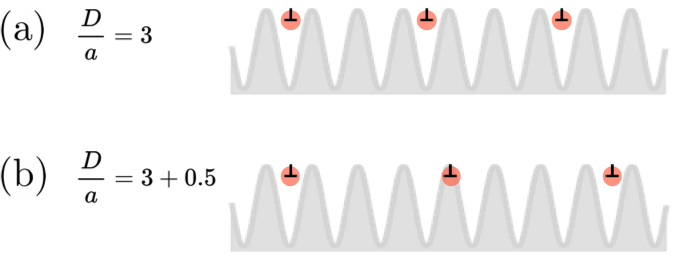}%
\caption{\setstretch{1.} Illustrations of a pileup commensurate with the underlying Peierls potential (gray) with $D/a = 3$ (a) and a pileup incommensurate with the underlying Peierls potential with $D/a = 3.5$ (b), where $a$ is the host lattice constant and $D$ is the average dislocation spacing. \label{fig:dp}}
\end{figure}

The Hamiltonian for a more general incommensurate pileup in real space is 
\begin{eqnarray} \label{eq:H_rough_dp}
H &=&  \left( \frac{Y b^2}{8 \pi } \right)  \sum_{n\neq m} \ln |(R_n - R_m) + (u_n - u_m)|\\
&& - V_\text{Peierls} \sum_n    \cos \left( \frac{ 2 \pi (R_n + u_n) }{a} \right), \notag
\end{eqnarray}
where the equilibrium positions of the dislocations $R_n$  can be decomposed as $R_n= nD = na (M + c)$, with
\begin{eqnarray}
\frac{D}{a} = M + c,
\end{eqnarray}
where $M = \lfloor \frac{D}{a} \rfloor$ is the maximum integer less than $\frac{D}{a}$ and $c$ is the non-integral part of the decomposition (we assume for simplicity a constant dislocation spacing $D$ here). When $c = 0$, $R_n$ drops out of the second term in Eq.~(\ref{eq:H_rough_dp}) and the Hamiltonian reduces to that of the accidentally commensurate pileup in Eq.~(\ref{eq:H_P}). 
The transition temperature $T_P^0$ for a commensurate pileup $c=0$ in Eq.~(\ref{eq:TF0}) can then be written as
\begin{eqnarray} \label{eq:TF02}
k_B T_P^0 =\frac{2 }{M^2} \frac{Yb^2}{16 \pi} . 
\end{eqnarray}

In contrast, when $c \neq 0$ (an illustration of
a pileup with nonzero incommensurability is shown in Fig.~\ref{fig:dp}b), $R_n$ cannot be neglected in the Peierls potential term and a quantitative treatment of the Hamiltonian in Eq.~(\ref{eq:H_rough_dp}) is more difficult.
There may still be a pinning transition at low temperatures, since a strong incommensurate periodic potential can pull the floating defect into registry with the host lattice despite the long range interactions. To see that the accidentally commensurate pinning temperature $T_P^0$ is an upper bound on the pinning temperature $T_P$ for nearby commensurate densities, note that incommensurate perturbations to the floating solid should be \textit{less} relevant than the commensurate perturbations we considered above, similar to two-dimensional commensurate-incommensurate transitions with short range interactions~\cite{nelson1979dislocation,halperin1978theory,nelson2002defects}. Physically, this means that the effect of an incommensurate periodic potential on a pileup is even more likely to be averaged out at long wavelengths than that of a commensurate periodic potential. For a system of particles with short-range interactions subject to a periodic potential, one can see explicitly how the pinning temperature decreases as a function of increasing incommensurability, by studying the effect of a perturbing pressure on the particles of an accidentally commensurate lattice~\cite{nelson1979dislocation,halperin1978theory,nelson2002defects}. In the case of our defect lattice, \textit{long-range} interactions between the dislocations make the pileup incompressible, which further restricts the possible accommodations (e.g. domain wall profiles) to the periodic potential at long wavelengths. Hence, the temperature $T_P$ of the transition from the floating-defect state to the pinned-defect state for pileups with nonzero incommensurability $c$ is depressed from that of accidentally commensurate pileups. More precisely, we expect that the pinning temperatures $T_P$ of a pileup with average dislocation space $D = a(M + c)$, where $c \neq 0$, is bounded from above by the temperature $T_P^0$ in Eq.~(\ref{eq:TF02}),
\begin{eqnarray} \label{eq:TF}
T_P \leq T_P^0.
\end{eqnarray}We cannot rule out the possibility that $T_P$ is depressed all the way to $T = 0$ for some incommensurate dislocation densities, due to long range interactions. 

\subsection{Structure factor}

When a pileup is in the pinned-defect phase at, say, an accidentally commensurate density, the displacement of every dislocation is close to zero $u_n \approx 0$. Consider for concreteness an accidentally commensurate pileup in Eq.~(\ref{eq:H_rough_dp}) with $R_n = nMa$. Since the phonon displacements $\{ u_n \}$ are small, we can expand the cosine Peierls potential term up to quadratic order in the displacements $u_n$. Upon discarding a constant term, we obtain the Hamiltonian for a uniform pileup in the pinned-defect phase
\begin{eqnarray} 
H
&=& \frac{1}{2} \frac{Y b^2}{4 \pi D^2} \int \frac{dq}{2 \pi} |q| |u (q)|^2 -  \frac{ 2\pi^2  }{a^2}  V_\text{Peierls} \sum_n   u_n^2. \notag
\end{eqnarray}
The displacement correlation function $C(s)$ in Eq.~(\ref{eq:Gs_def}) then approaches a finite value for large separation distances $s \rightarrow \infty$. It is straightforward to show that the effect of thermal fluctuations near the $m$-th reciprocal lattice vector $q \approx G_m$ leads to the following structure factor $S(q)$,
\begin{eqnarray}
\lim_{q \rightarrow G_m} S(q)  &=&  \sum_{s = -\infty}^{ \infty} e^{i(q - G_m) Ds} e^{-\frac{1}{2} G_m^2 \langle |u_s - u_0|^2 \rangle} \\
&=& \frac{2 \pi}{D}\delta(q - G_m) e^{- m^2 \frac{a^2}{D^2}\frac{k_B T}{V_\text{Peierls}}}. 
\end{eqnarray}
Thus, in the pinned-defect phase, the structure factors of pileups exhibits delta function Bragg peaks at all reciprocal lattice vectors $\{ G_m\}$, each suppressed by a Debye-Waller factor~\cite{ashcroft1976solid}. The corresponding oscillations in the radial distribution function $g(r)$ no longer decay for large separation distances $r$. 

The pinning transition from the floating defect phase may be observed through the structure factor as follows. Just above the pinning transition, i.e. as $T \rightarrow T_P^+$, pileups in the floating-defect phase will display a set of algebraically diverging Bragg peaks at the lower order reciprocal lattice vectors $G_m$, with
\begin{eqnarray}
|m| < \sqrt{\frac{T_C^{(1)}}{T_P}}. 
\end{eqnarray}
Below the transition $T=T_P$, these algebraically diverging Bragg peaks will narrow into delta function Bragg peaks. In contrast, at the same transition point, delta function Bragg peaks rise up from originally \textit{non-diverging} Bragg peaks at the higher order reciprocal lattice vectors $G_m$, with 
\begin{eqnarray}
|m| > \sqrt{\frac{T_C^{(1)}}{T_P}}. 
\end{eqnarray}

\subsection{Inhomogeneous pileups}
For pileups whose dislocation spacings are slowly varying as a function of space $D \rightarrow D(x)$, the temperature of the transition from the floating-defect phase to the pinned-defect phase varies locally throughout the pileup, i.e. $T_P \rightarrow T_P(x)$. Specifically, the local transition temperature for an accidentally commensurate region of an inhomogeneous pileup becomes
\begin{eqnarray} \label{eq:TF0}
k_B T_P^0(x) = \frac{Yb^2}{16 \pi} \frac{2 a^2}{D(x)^2},
\end{eqnarray}
which shifts the entire curve of transition temperatures $T_P(x)$ for incommensurate inhomogeneous pileups according to Eq.~(\ref{eq:TF}). Thus, upon decreasing the temperature, the dislocations in the denser regions of an inhomogeneous pileup with smaller dislocation spacing $D(x)$ will get pinned before the dislocations in the sparser regions. For example, in the semicircle pileup examined in the last section, the dislocations near the center of the pileup will be pinned by Peierls potential starting at higher temperatures compared to the more dilute dislocations near the edges of the pileup. One could probe this gradual regional pinning by measuring the local radial distribution function, or identifying the narrowing of the Bragg peaks in the spatially-averaged structure factor, which is now a combination of delta functions, from regions where the defects are pinned, and power laws, from regions where the defects are floating. 

\section{Conclusions \label{sec:conc}}
We have used statistical mechanics to explore a series of one-dimensional melting phase transitions in floating defect phases, as well as a low-temperature transition from a floating-defect phase to a pinned defect phase of 1d dislocation pileups embedded in 2d crystals. Importantly, all results in this paper can be experimentally probed through the structure factor and spatial correlation functions. 

The long-range interactions within one-dimensional dislocation assemblies play a crucial role in the physics of pileups. First, they describe the quasi-long range ordered defect-solid phase, for which we have derived a set of temperature-dependent critical exponents, which control power law divergences of the Bragg peaks in the structure factor $S(q)$ near the reciprocal lattice vectors $G_m$ and the algebraic decay of correlations in the radial distribution function $g(r)$. The sequential melting of pileups to more disordered defect-liquids is characterized by the serial disappearance of algebraically diverging Bragg peaks at temperatures given by Eq.~(\ref{eq:Tcm}). We have obtained the exact forms of the structure factor, for both uniform and inhomogeneous pileups, close to $q = 0$ and to the reciprocal lattice vectors $q = G_m$, and found agreement with random matrix simulations. In addition, these long range repulsive interactions can also conspire with Peierls potential to pin down the defects at low temperatures, facilitating the transition between quasi-long range ordered floating defects to the long range ordered pinned defects, characterized by the transformation of power law Bragg peaks to delta function Bragg peaks at temperatures given by Eqs.~(\ref{eq:TF}) and (\ref{eq:TF0}). 

By extending our investigations to pileups with non-uniform densities, we have also explored the statistical mechanics of a new class of one-dimensional inhomogeneous defect crystals. By mappings to random matrix theory and the quantum Brownian motion model, we have obtained the phase boundaries, structure factors, and local pair correlation functions for a variety of one-dimensional phases exhibited by dislocation pileups. 
Interestingly, the sequence of melting temperatures associated with pileup structure factors embody a collective behavior independent of their local density. In contrast, the low temperature pinning transition depends strongly on the ratio between the local average dislocation spacing and the host lattice constant, leading to a series of localized pinning transitions in inhomogeneous pileups. 

Dislocation pileups in two-dimensional crystals have been observed in previous simulations of spherical crystalline caps (see, for example, Fig.~3f of Ref.~\cite{azadi2014emergent} and Fig.~5a of Ref.~\cite{azadi2016neutral}). 
Significantly, disclinations and grain boundary scars are required as part of the ground state for certain curved crystals by topological constraints~\cite{bowick2000interacting} and have been observed experimentally in a variety of systems, including spherical colloidosomes~\cite{bausch2003grain,irvine2010pleats} and bubble rafts on the surface of rotating fluids~\cite{bowick2008bubble}. 
Such systems, in addition to liquid crystal thin films, electrons on the surface of helium, and others~\cite{strandburg1988two}, on both flat and curved surfaces, are among the possible experimental platforms that can be used to investigate the physics of dislocation pileups described in this work. 

In three dimensional bulk materials, multiple pileups can emanate from a single stress source and interact with each other on the same glide plane (see Fig.~\ref{fig:intro_schm}a). In the limit of short, rigid defect lines, the statistical mechanics reduces to the two-dimensional case we've considered here. However, the statistical mechanics of more general longer and deformable dislocation lines could be studied by mapping to interacting bosonic worldlines in quantum mechanical path integrals~\cite{nelson2002defects}. Here, long range interactions along the time-like direction could be important~\cite{moretti2004depinning}. Previous works have studied the dynamics and scaling morphologies of dislocation lines in disordered stress landscapes and as tangled cell structures~\cite{sethna2017deformation}. It would be interesting to investigate the effects of thermal fluctuations on these even more complex dislocation structures, and on the pinning effects of Peierls potential in three dimensional dislocation assemblies on the problems studied in Ref.~\cite{moretti2004depinning}. 

\begin{acknowledgements}
It is a pleasure to acknowledge helpful conversations with F. Spaepen, I. Garcia-Aguilar, and J. Huang. G.H.Z. acknowledges support by the National Science Foundation Graduate Research Fellowship under Grant No. DGE1745303. This work was also supported by the Harvard Materials Science and Engineering Center, via National Science Foundation Grant No. DMR-2011754 as well as through NSF Grant No. DMR-1608501.
\end{acknowledgements}
\appendix

\section{Structure factor of a 1d lattice with short range interactions \label{app:SR}}

In this section, we review the structure factor of a uniform one-dimensional array of particles with short range interactions~\cite{kardar2007statistical,emery1978one}.  We work in the classical limit, and assume that the momentum degrees of freedom have been integrated out. 
Consider an infinitely long homogeneous 1d lattice, whose particle density is given by
\begin{eqnarray}\label{eq:dens}
\rho(x) = \sum_{n = - \infty}^\infty \delta (x - x_n)
\end{eqnarray}
where $n$ sums over all particles, and $x_n = nD + u_n$ is the instantaneous position of the $n$-th particle and $u_n$ is the displacement of the $n$-th particle from its equilibrium position due to fluctuations. If the particles interact with their nearest neighbors via a potential energy of the form \small
\begin{eqnarray}
E = \frac{K}{2} \sum_n (x_{n+1}-x_n - D)^2 =  \frac{K}{2} \sum_n (u_{n+1}-u_n)^2,
\end{eqnarray} \normalsize
then the energy in the continuum limit is given by \small
\begin{eqnarray}
\Delta E = \frac{B_0}{2}\int dx \left( \frac{du(x)}{dx} \right)^2 = \frac{B_0}{2}  \int \frac{dq}{2 \pi} q^2  |u(q)|^2,
\end{eqnarray} \normalsize
where $B_0 = \frac{K}{D}$ is a constant. 
The Fourier coefficients of the density in Eq.~(\ref{eq:dens}),
\begin{eqnarray}
\rho(q) = \int dx e^{i q x} \sum_n \delta(x-x_n) = \sum_n e^{i q x_n},
\end{eqnarray}
determine the structure factor via 
\begin{eqnarray} \label{eq:Sq_s}
S(q) =\langle | \rho(q) |^2 \rangle = \sum_{s = - \infty}^{\infty} e^{i q sD} \left \langle e^{iq(u_s - u_0)} \right \rangle,
\end{eqnarray}
where $u_s = u(x = sD)$. 
At low temperatures, we expect that $S(q)$ is large near the reciprocal lattice vectors $q \approx G_m = \frac{2 \pi}{D} m$, where $m = 0, \pm 1, \cdots$. We now expand about the $m$-th reciprocal lattice vector $G_m$ by setting $q = G_m + k$ with $k \ll G_m$, and approximate Eq.~(\ref{eq:Sq_s}) as
\begin{eqnarray} \label{eq:Sq_approx}
S(q) \approx \sum_{s = - \infty}^{\infty} e^{i k sD} \exp\left[  - \frac{1}{2} G_m^2 \left \langle (u_s - u_0)^2  \right \rangle\right] . 
\end{eqnarray}
The displacement correlation function needed in Eq.~(\ref{eq:Sq_approx}) is \small
\begin{eqnarray}
C(s) &\equiv& \left \langle (u_s - u_0)^2  \right \rangle = \int_{-\pi/a}^{\pi/a} \frac{dp}{2 \pi} \frac{k_B T}{B_0p^2} \left( 1- e^{i psD} \right)  \label{eq:Gs_SR}\\
&\approx & \frac{k_B T Ds}{\pi B_0} \int_0^\infty d \eta \frac{ 1 - \cos (\eta)}{\eta^2}  = \frac{k_B TDs}{2 B_0},
\end{eqnarray} \normalsize
where we can extend the integration limits in Eq.~(\ref{eq:Gs_SR}) to $\pm \infty$ because the integral is dominated by small wavevectors $p$. 
Upon substituting this displacement correlation into Eq.~(\ref{eq:Sq_approx}), we have
\begin{eqnarray}
S(q = G_m + k) \approx \frac{2}{D} \int_0^\infty  dx \cos (kx) e^{- \frac{k_B T}{4 B_0} G_m^2 |x|}. 
\end{eqnarray}
Since
\begin{eqnarray}
\int_0^\infty dx e^{- \alpha x} \cos (kx) = \frac{\alpha}{\alpha^2 + k^2},
\end{eqnarray}
we obtain the structure factor as a sum of Lorenzians centered at each reciprocal lattice vector $G_m$
\begin{eqnarray}
S(q) \approx \frac{2}{D} \sum_{m \neq 0} \frac{\xi_m^{-1}}{ \xi_m^{-2} + (q - G_m)^2},
\end{eqnarray}
where the widths of the Lorentzians $\{ \kappa_m \equiv \xi^{-1}_m \}$ are given by
\begin{eqnarray}
\xi_m = \frac{ 4 B_0}{ k_B T G_m^2}. 
\end{eqnarray}
Thus, we have a set of correlation lengths, one associated with each Lorentzian peak. Note that these peaks become less pronounced with increasing $m$ and that the correlation lengths $\xi_m \sim 1/m^2 k_B T$ only diverge in the limit $T \rightarrow 0$. 

\section{Equilibrium density of dislocation pileups} \label{app:DP}

The average dislocation density $n(x)$ resulting from a particular applied shear stress profile $\sigma(x)$ can be obtained by solving the force balance condition in Eq.~(\ref{eq:force_balance}). In this section, we derive the average dislocation densities for the different pileups shown in Table~\ref{tab:disloc} using the continuum framework explained in Sec.~\ref{sec:DP}. 

Throughout these calculations, we will frequently employ the Hilbert transform $H_x [ \cdot]$, defined by~\cite{bracewell1986fourier} 
\begin{eqnarray}
H_x[f(y)] = \frac{1}{\pi} \text{PV} \int_{-1}^{1} \frac{f(y)}{y-x} dy,
\end{eqnarray}
where PV denotes the principal part of the improper integral. 
Special solution pairs to the Hilbert transform are given by the Tschebyscheff (Chebyshev) polynomials $T_n$ and $U_n$, with $f(y) \equiv T_n(y)/\sqrt{1 -y^2}$ or $f(y) \equiv U_{n-1}(y) \sqrt{1 - y^2}$,
\begin{eqnarray}
H_x \left[ \frac{T_n(y)}{\sqrt{1 - y^2}} \right] = U_{n-1}(y) \label{eq:cheby}\\
H_x \left[ U_{n-1}(y)\sqrt{1 - y^2} \right] = -T_{n}(y).  
\end{eqnarray}

\subsection{Spatially uniform stress fields: Double and single pileups} \label{sec:DP_uni}
When the stress field is uniform in space and the stress source is located at the center between the dislocation barriers, a double dislocation pileup occurs via the Frank-Read mechanism~\cite{frank1950multiplication} (see row 1 of Table~\ref{tab:disloc}). 

In the continuum model, the Hamiltonian for the double dislocation pileup is
\normalsize   
\begin{eqnarray} \label{eq:H_cont_D}
H [ n ( x ) && ]= - \int_{-L/2}^{L/2} d x ~n ( x ) \sigma_0 b x  \\
- \frac{1}{2}&& \frac { Y b ^ { 2 } } { 4 \pi} \int_{-L/2}^{L/2} dx \int_{-L/2}^{L/2} dx^\prime n ( x ) n \left( x ^ { \prime } \right) \ln{| x - x ^ { \prime } |} , \notag
\end{eqnarray} \normalsize  
where $n(x)$ is the density of dislocations along the pileup, which can be positive, referring to the density of dislocations with positive Burgers vector $\vec b = b \hat x$, or negative, referring to the density of dislocations with negative Burgers vector $\vec b = -b \hat x$. 

The force balance condition at equilibrium is then~\cite{hirth1983theory}
\begin{eqnarray}
  \sigma_0 b = \frac { Y b ^ { 2 } } { 4 \pi}\int_{-L/2}^{L/2} dx^\prime \frac{ n \left( x ^ { \prime } \right)}{x^{ \prime } - x }.
\end{eqnarray}
Note that the 2d Young's modulus has dimensions of force per unit length, $[ Y] = \frac{F}{L}$. 

We obtain the equilibrium dislocation density using the Chebyshev polynomials $U_0(x)$ and $T_1(x)$, as shown in Eq.~(\ref{eq:cheby}),
\begin{eqnarray}
n_\text{D}(x) = \zeta \frac{x}{\sqrt{\left(\frac{L}{2}\right)^2-x^2}}, 
\end{eqnarray} where $\zeta = \frac{4 \sigma_0}{Y b}$ and $x \in \left (-\frac{L}{2}, \frac{L}{2} \right) $~\cite{hirth1983theory}. Note that $n_\text{D}(x)$ has different signs for $x >0$ and $x<0$, meaning the dislocations on opposite sides of the pileup have oppositely oriented Burgers vectors: $\vec b = \text{sign}(x) b \hat x$ (see schematic in Table~\ref{tab:disloc}). The normalization condition $\int_{-L/2}^{L/2} dx n_\text{D}(x) = N_\text{D}$ gives the expression for the total number of dislocations in a double pileup $N_\text{D}$ as
\begin{eqnarray}
 N_D = \zeta L. 
\end{eqnarray}
Note that $\zeta$ has dimensions of inverse length, $ \left[ \zeta \right] = L^{-1}$, so that $N_D$ is dimensionless, as expected.

To account for the single (stressed) pileup (row 2 of Table~\ref{tab:disloc}), we need to use the \textit{unstressed} single pileup density $f(x)$,
\begin{eqnarray}
 f(x) = \frac{1}{\sqrt{1 - \left( \frac{x}{L/2} \right)^2 }},
\end{eqnarray}
derived via the following force balance condition
\begin{eqnarray}
0 = \frac{1}{\pi} \text{PV}\int_{-L/2}^{L/2} \frac{f (x')}{x'- x} d x'. 
\end{eqnarray}
The density distribution of a single \textit{stressed} dislocation pileup $n_\text{S}(x)$ is then obtained from a linear combination of the double pileup density $n_\text{D}(x)$ and the \textit{unstressed} single pileup density $f(x)$, with coefficients $a'$ and $b'$ such that the normalization condition is satisfied and the dislocation density vanishes at the right side of the pileup $x = L/2$, \small  
\begin{eqnarray} \label{eq:nS}
n_\text{S}(x=L/2) = a' f(x=L/2) + b' n_\text{D}(x=L/2) = 0.
\end{eqnarray} \normalsize  
The density distribution for the single stressed dislocation pileup (centered at $x=0$) is then
\begin{eqnarray} \label{eq:n_single}
n_\text{S}(x) = \zeta\frac{\sqrt{L/2 - x}}{\sqrt{L/2 + x}},
\end{eqnarray}
where $x \in \left (-\frac{L}{2}, \frac{L}{2} \right) $. 
Note that the total length of the pileup is significantly impacted by the last few, widely spaced, dislocations on the right side of the pileup, which are poorly represented in the continuum model. Nevertheless, Eq.~(\ref{eq:n_single}) accurately describes most of the pileup distribution~\cite{hirth1983theory}. The normalization condition relating the length and the total dislocation number in a single stressed pileup is given by
\begin{eqnarray}
N_\text{S} &=& \zeta\frac{L}{2} \pi. 
\end{eqnarray}

\subsection{Spatially varying stress fields: semicircle lattice and uniform lattice} \label{sec:DP_var}

If the shear stress varies linearly in space and, in particular, smoothly changes sign at the center of the interval, the dislocations in the pileup then form a semicircular lattice (row 3 of Table~\ref{tab:disloc}). 

In the continuum limit, the Hamiltonian for the semicirclular dislocation lattice is
\begin{eqnarray} \label{eq:H_cont}
H [ n ( x ) &&] = \int_{-L/2}^{L/2} d x n ( x ) \frac{\sigma_0 b }{L/2} \frac{x^2}{2} \\
-\frac{1}{2}&& \frac { Y b ^ { 2 } } { 4 \pi} \int_{-L/2}^{L/2} dx \int_{-L/2}^{L/2} dx^\prime n ( x ) n \left( x ^ { \prime } \right) \ln{| x - x ^ { \prime } |}, \notag
\end{eqnarray}
where the factor of $L/2$ in the denominator of the first term makes manifest that $[ \sigma_0 b \frac{x}{L} ] = F$ has dimensions of force. 
The force balance condition that results is then
\small 
\begin{eqnarray}
\sigma_0 b \frac{x}{L/2} = \frac { Y b ^ { 2 } } { 4 \pi} \text{PV} \int_{-L/2}^{L/2} dx^\prime \frac{ n \left( x ^ { \prime } \right)}{x' - x}.
\end{eqnarray}\normalsize 
Using the relevant pair of Chebyshev polynomials, $T_1(x) = x$ and $U_0 = 1$, we obtain the dislocation density as
\begin{eqnarray}
n_\text{SC}(x) = \zeta \sqrt{1 - \left(\frac{x }{L/2}\right)^2},
\end{eqnarray}
where, as before, $\zeta = 4 \sigma_0/Yb$ and $x \in \left (-\frac{L}{2}, \frac{L}{2} \right)$. The condition $\int_{-L/2}^{L/2} dx~ n_\text{SC}(x) = N_\text{SC} $ relates $\zeta$ to the total number of dislocations as
\begin{eqnarray}
N_\text{SC} = \zeta \frac{L \pi}{4}. 
\end{eqnarray}

One can also use the force balance condition to invert for the stress field corresponding to a \textit{uniform} dislocation lattice (row 4 of Table~\ref{tab:disloc}). We start by first solving for the applied force profile from the force balance condition in Eq.~\ref{eq:force_balance}. Upon letting $n_\text{U}(x)$ denote a prescribed \textit{constant} density of dislocations in a pileup, described by a rectangle function $\Pi(z)$,
\begin{eqnarray}
n_U(x) = n_\text{U}\Pi\left( \frac{x}{L} \right) \equiv \begin{cases} 1 & \left| \frac{x}{L} \right| < \frac{1}{2} \\ 
\frac{1}{2} & \left| \frac{x}{L} \right| = \frac{1}{2}  \\
0 & \left| \frac{x}{L} \right| > \frac{1}{2}  \end{cases},
\end{eqnarray}
the force balance condition for the uniform pileup spanning $x \in ( - \frac{L}{2}, \frac{L}{2})$ is
\begin{eqnarray}  \label{eq:UUx}
-\sigma_0 b U'_\text{U}(x) &=& \frac { Y b ^ { 2 } } { 4 \pi} \int_{-L/2}^{L/2}\frac{n_U}{y-x} dy.
\end{eqnarray}
Since the rectangle function $\Pi(z)$ has the following well-defined Hilbert transform~\cite{bracewell1986fourier},
\begin{eqnarray}
\mathcal{H}[\Pi(z)]&=&\frac{1}{\pi} \text{PV} \int_{-\infty}^{\infty} \frac{\Pi(z) d z}{z-y} \\
&=&\frac{1}{\pi} \text{PV} \int_{-1/2}^{1/2} \frac{\Pi(z) d z}{z-y} = \frac{1}{\pi} \ln \left|\frac{y-\frac{1}{2}}{y+\frac{1}{2}}\right|, 
\end{eqnarray}
Eq.~\ref{eq:UUx} becomes
\begin{eqnarray}
U'_\text{U}(x) &=& -\frac{n_\text{U}}{\zeta \pi }  \ln \left|\frac{x-\frac{L}{2}}{x+\frac{L}{2}} \right|. \label{eq:F_uni}
\end{eqnarray}
Upon integrating Eq.~(\ref{eq:F_uni}), the central potential that gives a constant density of dislocations in a pileup is obtained as 
\begin{eqnarray} \label{eq:U_uni}
U_\text{U}(x) &=& \left( \frac{L}{2} - x \right) \ln \left(  \frac{L}{2} - x \right) \\
&&+  \left( \frac{L}{2} + x \right) \ln \left( \frac{L}{2} + x \right) \notag
\end{eqnarray}
(see row 4 of Table~\ref{tab:disloc}) and the uniform dislocation density associated with the potential is given by
\begin{eqnarray}
n_\text{U} =  \zeta\pi. 
\end{eqnarray}
The normalization condition, in this case straightforward, gives
\begin{equation}
N_\text{U} = \zeta\pi L = \frac{4 \sigma_0}{Yb} \pi L. 
\end{equation}

It is interesting to compare the central potential profiles for the semicircular lattice $U_\text{SC}(x)$ and the uniform lattice $U_\text{U}(x)$. To this end, we set $ x = \frac{L}{2} y$, such that the pileups lie on the interval $y \in [-1, 1]$. The discrete Hamiltonians for the semicircular lattice and the uniform lattice then read
\small  
\begin{eqnarray}
H_\text{SC} 
&=&  \sigma_0^\text{(SC)} b \frac{L_\text{SC}}{4} \left[ \sum_i y_i^2 - \frac{1}{2 N_\text{SC}} \sum_{i \neq j} \ln | y_i - y_j|\right]   \label{eq:HSC_imp}\\
H_\text{U} 
&=&  \sigma_0^\text{(U)} b L_\text{U}\Bigg[ \frac{1}{2}\sum_i \left (1 -y_i \right) \log \left (1 - y_i \right) + \left (y_i+1 \right)\log \left (y_i+ 1 \right)    \notag \\ 
&& \quad \quad  - \frac{1}{2 N_\text{U}} \sum_{i \neq j} \ln | y_i - y_j|\Bigg]. \label{eq:HU_imp}
\end{eqnarray} \normalsize  
To appropriately compare the central potential profiles for the semicircle lattice and the uniform lattice, we set the total number of dislocations in both pileups to be equal
\begin{eqnarray} \label{eq:N_SC_U}
N_\text{SC} &=& N_\text{U} \equiv N,
\end{eqnarray}
which, using the normalization conditions $N_\text{SC} = \zeta_\text{SC} L_\text{SC}$ and $N_\text{U} = \zeta_\text{U} L_\text{U} \pi$ (see third column of Table~\ref{tab:disloc}), translates to
\begin{eqnarray}
\frac{\sigma_0^\text{(SC)} L_\text{SC}}{4} = \sigma_0^\text{(U)} L_\text{U} \equiv \sigma_0 L.  \label{eq:sigmaL_SC_U}
\end{eqnarray}
Upon substituting in Eqs.~(\ref{eq:N_SC_U}) and (\ref{eq:sigmaL_SC_U}) into the discrete Hamiltonians in Eqs.~(\ref{eq:HSC_imp}) and (\ref{eq:HU_imp}) and expanding the central potential of the uniform pileup to quadratic order in $y$, we obtain
\small  
\begin{eqnarray}
H_\text{U} (\{y_i \approx 0 \})&=& H_\text{SC} (\{y_i \})  \label{eq: HSC_HU} \\
&=&  \sigma_0 b L \left[ \sum_i y_i^2 - \frac{1}{2 N} \sum_{i \neq j} \ln | y_i - y_j|\right].  
\end{eqnarray} \normalsize  
Thus, the central potential for the uniform lattice in Eq.~(\ref{eq:HU_imp}), when expanded around $y = 0$, gives the same factor of $y^2$ as the central potential for the semicircular pileup, and the semicircular and uniform dislocation pileups should have nearly identical statistical mechanics near the pileup center. 

\bibliography{references.bib}
\end{document}